\documentclass[%
 reprint,
 amsmath,amssymb,
pra,
]{revtex4-2}

\usepackage{graphicx}
\usepackage{stfloats}
\usepackage{dcolumn}
\usepackage{bm}

\usepackage{multirow}
\usepackage{graphics}
\usepackage{xcolor}
\usepackage{amssymb}
\usepackage{braket}
\usepackage{longtable}
\usepackage{rotating}
\usepackage{diagbox}
\usepackage{enumitem}
\usepackage{nicefrac}

\usepackage{tikz}
\usetikzlibrary{tikzmark}

\usepackage{changes}

\DeclareMathOperator{\atantwo}{atan2}

\usepackage{scalerel}
\usepackage[normalem]{ulem}

\newcommand{\stkout}[1]{\ifmmode\text{\sout{\ensuremath{#1}}}\else\sout{#1}\fi}

\newcommand{\ME}[3]{ \mbox{$\langle #1\,|\,#2\,|\,#3\rangle $} }
\newcommand{\MEred}[3]{ \mbox{$\langle #1\,||\,#2\,||\,#3\rangle $} }
\newcommand{\CGC}[6]{ \mbox{$ ( #1 #2 , #3 #4\,|\, #5 #6 ) $} }
\newcommand{\SechsJ}[6]{ \mbox{$ %
            \arraycolsep0.25ex %
            \left\{ \begin{array}{ccc} %
                       #1 & #2 & #3 \vspace{0.5ex}\\%
                       #4 & #5 & #6 %
                   \end{array} \right\} $} }

\begin{document}

\preprint{APS/123-QED}

\title{General properties of the RABBITT at parity mixing conditions}

\author{Maria M. Popova}
 \affiliation{ 
Skobeltsyn Institute of Nuclear Physics, Lomonosov Moscow State University, 119991 Moscow, Russia
}
 \affiliation{ A.V. Gaponov-Grekhov Institute of Applied Physics, Russian Academy of Sciences, 603950 Nizhny Novgorod, Russia}
 \affiliation{ 
MIREA --- Russian Technological University, Moscow, 119454, Russia
}

\author{Sergei N. Yudin}
 \affiliation{ 
Skobeltsyn Institute of Nuclear Physics, Lomonosov Moscow State University, 119991 Moscow, Russia
}

\author{Alexei N. Grum-Grzhimailo}
\affiliation{ 
Skobeltsyn Institute of Nuclear Physics, Lomonosov Moscow State University, 119991 Moscow, Russia
}

\author{Elena V. Gryzlova}
\affiliation{ 
Skobeltsyn Institute of Nuclear Physics, Lomonosov Moscow State University, 119991 Moscow, Russia
}
\email{gryzlova@gmail.com}

\date{\today}

\begin{abstract}
Parity mixing in photo\-ionization, i.e.  when emitted electrons have different parities but the same energy, causes interference observable only in  angle-resolved  measurements. The interference typically manifests  as a symmetry violation in the photo\-electron angular distributions.
The traditional, based on HHG, RABBITT scheme with high-order harmonics separated by twice the seed field energy, precludes parity mixing. On the contrary, a free-electron laser  provides a possibility to generate even harmonics. Using triple the fundamental frequency as a seed, one obtains a comb of alternating even and odd harmonics, separated by three times the initial frequency \cite{Maroju2020} (2-SB RABBITT). In this setup, there are two sidebands between the main photo\-electron lines, versus one in the traditional scheme. 
In the paper, we examine the general properties of a two-sideband scheme and analyze the symmetry breakdown of photo\-electron angular distributions for various polarization geometries of the incident pulse. We found a crucial difference in symmetries between 2-SB RABBITT and other photo\-ionization schemes with parity mixing. Illustrative calculations are carried out for neon with pulse parameters typical for modern facilities. The possibility to reconstruct the temporal profile of the pulse from the angle-resolved measurements is discussed. 
\end{abstract}


\maketitle


\section{\label{sec:1} Introduction}

The ability to control the polarization of light helps shed light on the process of light-matter interactions, such as photo\-excitation and photo\-ionization. Some polarization effects, such as circular magnetic dichroism, manifest themselves in angle-integrated spectra \cite{Burke,Fano}. Others, such as production of polarized photoelectrons, demand spin- and angle-resolved measurements, which strongly limits observation capabilities. The development of extreme ultraviolet (XUV) and X-ray (to limit to few-photon processes) bright coherent light sources, such as high-order harmonic generation (HHG) setups \cite{Lewenstein1994,strelkov2016} or X-ray free electron lasers (XFELs) \cite{CALLEGARI20211}, made angle-resolved experiments possible. 
Generating X-ray radiation with arbitrary polarization has long been a challenge.
While nonlinear polarization is achieved on XFELs and available for user operation \cite{Allaria2014,Schmising2017}, generating HHG with arbitrary polarization is not well-established yet. Up-to-date, a few methods have been proposed \cite{Kfir2015,Mahieu2018,Khokhlova2021,Emelin}.

Parity mixing, i.e. interference between channels of different parity, is another interesting phenomenon that can be observed only in angle-resolved measurements. Leaving aside the `uncontrolled' situation when parity mixing is a result of the dipole approximation violation,  the simplest showcase is the  {ionization in the bichromatic field that consists of the fundamental $\omega$ and second harmonic $2\omega$} ---  {the} `$\omega+2\omega$' scheme (in  {the} XUV regime, \cite{grum2015,NATPHOT-Prince-2016}, in  {the} strong-field ionization (SFI) regime, \cite{Frolov2010,Mancuso2016}). It has been shown that the polarization of light can drastically change observables in the presence of parity mixing \cite{douguet2016,Gryzlova19,Jasarevic2023,Gebre2024}.

Dynamic aspects of photo\-ionization are successfully studied using techniques such as streaking spectroscopy \cite{Itatani2002}, in which an atom is ionized by an XUV photon{,} and then  {the} freed electron in the continuum is subsequently  driven by the dressing laser field, and the RABBITT scheme \cite{Veniard1996,Paul2001}, in which an electron is promoted to the continuum by an XUV harmonic forming a mainline (ML) and then additionally absorbs or emits an optical (infrared --- IR) photon to form a sideband (SB). These techniques unlocked atto\-second time scales in experimental physics \cite{smirnova2009high, moioli2025}. A few other techniques have been developed, and some of them involve parity mixing \cite{you2020}.

In the traditional RABBITT scheme, the XUV harmonics differ by  {double infrared field frequency} $\omega$, and parity mixing cannot be achieved, as the interference occurs either between two-photon transitions in sidebands or between one- and three-photon transitions in mainlines. As there is only one sideband between the mainlines, following  \cite{Bharti2021} we refer to it as 1-SB scheme. Various modifications of this scheme based on replacing seed frequency $\omega$ with $2\omega$, $3\omega$ or even a comb of $\omega+2\omega$  have been proposed. In an HHG setup   with double fundamental frequency $2\omega$ \cite{Bharti2021} there are three sidebands between adjusted XUV harmonics --- the so-called 3-SB scheme; therefore, the interfering amplitudes in the sidebands have the same parity, being either two three-photon or two-photon and four-photon ones. In the setup based or triple fundamental frequency $3\omega$ realized at FERMI \cite{Maroju2020}, where assisted harmonics are even and odd, there are two SBs between each subsequent ML pair (2-SB scheme). In this setup, two-photon and three-photon amplitudes interfere and parity mixing occurs in both SBs and MLs. Great prospects opened with  {the} bi\-circular $\omega+2\omega$ scheme \cite{Kfir2015,Fan,Barreau}.
Using a comb of $\omega+2\omega$ allows creation of a phase-meter in the parity-mixing 0-SB scheme \cite{villeneuve2017,Donsa2019}.

Combining polarization control with atto\-second metrology is the next milestone toward understanding photoprocesses. It can help to extract information about continuum-continuum couplings \cite{Bharti2021}, pave the way to control  spin polarization \cite{Barth13,Hartung16,Milocevic2016,Gryzlova2020_2} or assess  molecular chirality \cite{Goetz2019,atoms6040061}.

The investigations have shown that polarization acts differently for atto\-second schemes with \cite{Popova2022,Jasarevic2023,Gebre2024} and without \cite{Hockett2017,Boll2017,Kheifets2023,popova2025} parity mixing. Here we develop an analysis of polarization effects on the  photoelectron angular distributions in the 2-SB RABBITT scheme. 

 Usually for theoretical description of RABBITT experiments, TDSE calculations are used. They have proven their efficiency, and, in fact, became a golden standard in the field: numerous theoretical calculations are performed in this approach, to name a few in addition to the already mentioned \cite{Bray2018,ocello2025,Serov2026}, however, most of them  don’t allow thorough analysis and fitting without running the whole simulation again, and are, in fact, a kind of numerical experiment conducted on a supercomputer. In the paper, we use two  less costly and time-consuming methods, which allow analyzing the contributions of various channels and use experimental energies of the states, and investigate their applicability to the 2-SB RABBITT scheme. Applicability test for 1-SB RABBITT case by a comparison with MCHF TDSE was done in \cite{moioli2025}. One should note that these methods can't replace TDSE without a significant loss in quality if a great number of states is involved, for example, for describing processes close to the ionization threshold.

Unless otherwise specified, the atomic system of units is used.

\section{\label{sec:2}Theoretical basement for the RABBIT{T} description}

In this paper, we further extend the approaches based on combining the solution of  an analog of rate equations  for the amplitudes (amplitude coefficients method --- ACE) and time-dependent perturbation theory (PT) applied earlier for the 1-SB RABBITT scheme \cite{popova2023,popova2025} to the 2-SB scheme,
similar to  \cite{Maroju2020}. Thus, here we only briefly describe the methods, clearly indicating the differences  arising between the 1-SB RABBITT and the 2-SB RABBITT schemes.

The 2-SB RABBITT scheme uses the advantage of XFEL to generate even harmonics as well as odd ones. The electromagnetic field is presented as a sum of  {an} XUV comb consisting  {of} harmonics of order $N=3n$  generated at a $3\omega$ frequency (excluding $3\omega$ itself)   together with  {an} $\omega$ pulse: 
\begin{align}\label{eq:field}
\bm{E}(t)&=\Re\Big{[}\sum_{N\Lambda\lambda}E_{\rm{xuv}}c_\Lambda\bm{\epsilon}_{\Lambda} e^{-i(N\omega t+\phi_{N})}+\nonumber\\
&\hspace{90pt}E_{\rm{ir}}c_\lambda\bm{\epsilon}_{\lambda}e^{-i(\omega t+\phi)}\Big{]},
\end{align}
where $E_{\rm{xuv}}=E_{\rm{xuv}}^0\cos^2(\frac{2t}{\tau})$ and $E_{\rm{ir}}=E_{\rm{ir}}^0\cos^2(\frac{t}{\tau})$ 
are slowly varying envelopes, $E_{\rm{ir}}^0$ and $E_{\rm{xuv}}^0$ are strengths of the IR and XUV components, and $\tau$ determines the pulse duration; $\phi_{N}$ is Nth XUV component' {s} phase, $\phi$ is the phase of the IR pulse connected with the IR pulse delay $\tau_{\rm del}$ as  $\phi=\omega \tau_{\rm del}$. In accordance with experimental conditions{,} we consider  {the} IR pulse as twice as long as the XUV. The field polarization is determined by a decomposition  {with coefficients $c_\Lambda$, $c_\lambda$} over cyclic coordinate vectors $\bm{\epsilon}_{\lambda/\Lambda=1}=-(\bm{\epsilon}_x+i\bm{\epsilon}_y)/\sqrt{2}$, $\bm{\epsilon}_{\lambda/\Lambda=-1}=(\bm{\epsilon}_x-i\bm{\epsilon}_y)/\sqrt{2}$ and $\bm{\epsilon}_{\lambda/\Lambda=0}=\bm{\epsilon}_z$,  {and as $\Lambda/\lambda$ is connected to XUV/IR photon helicity,} the latter appears if a field propagates not along the quantization axis $z$. 

Following \cite{popova2025}, we use the $LS$-coupling scheme, so  {the} $n$th eigen\-function of the unperturbed Hamiltonian $\psi_{\alpha_n}(\varepsilon,{\bm r})$ depend {s} on the following quantum numbers: energy $\varepsilon$, core (ion) orbital momentum $L_c$, active electron angular momentum $l$, total angular momentum $\bm L=\bm L_c+\bm l${,} and its projection $M$. Here we assume that the electric dipole operator does not affect spin and  {that} a ground state of  {an} atom has total spin $S=0$.
A wave function of the system $\Psi({\bm r}, t)$ is expanded in the basis of eigen\-functions of the unperturbed Hamiltonian:
\begin{align} \label{eq:wf_decomp}
&\Psi({\bm r}, t) = \sum_{L_clLM} \hspace{-5pt}\Big{(} \sum_{n} \mathcal  U_{(L_cl)LM}(\varepsilon,t) \psi_{\alpha_n}{(\varepsilon,{\bm r})} e^{-i  \varepsilon t}\nonumber\\
&\hspace{50pt}+\int \hspace{-3pt}d\varepsilon \,   \mathcal  U_{(L_cl)LM}(\varepsilon,t)\psi_{\alpha_\varepsilon}{(\varepsilon,{\bm r})}
e^{-i  \varepsilon t} \Big{)}\,,
\end{align}
where $U_{(L_cl)LM}(\varepsilon,t)$ are complex expansion coefficients and $\alpha_n$, $\alpha_\varepsilon=\{L_c,l,L,M\}$ mean a set of  quantum numbers to characterize a particular state that belongs to  {a} discrete or continuum spectrum.

To describe the continuum states in (\ref{eq:wf_decomp}),  discretization was applied, i.e.{,} integration was replaced by summation with  {a} uniform energy step $d\varepsilon$.
Therefore, $|  \mathcal  U_{(L_cl)LM}(\varepsilon,t) |^2$ is the  probability   {density} of finding an electron within a neighborhood $d\varepsilon$ of the energy $\varepsilon$ at time $t$.  {It is known that transitions through continuum diverge as $1/(\varepsilon-\varepsilon')^2$ in length gauge and as $1/(\varepsilon-\varepsilon')$ in velocity gauge \cite{Gordon1929, Veniard1989,Trippenbach_1989,Korol_1994}.}  In order to suppress   {the} divergences, the velocity gauge was applied with vector potential $\bm{A}=-\int \bm{E}(t)dt$. Each part of its decomposition into cyclic coordinates can be presented as a sum of the XUV ${A}_{\rm xuv}(t)$ and IR ${A}^{u/d}_{\rm ir}(t)$ components.  The component ${A}^u_{\rm ir}(t)$ is associated with the absorption of an IR photon and behaves as $e^{-i(\omega t+\phi)}$, and ${A}^d_{\rm ir}(t)$ --- with the emission and behaves as $e^{i(\omega t+\phi)}$.

Then the system of differential equations for expansion coefficients:
\begin{eqnarray}
     \label{eq:diff}
 &&\hspace{-9pt}\frac{d   \mathcal  U_{(L_cl')L'M'}(\varepsilon',t)}{dt} =\nonumber\\
 &&\hspace{-9pt}-i  \sum_{n} e^{i(\varepsilon'-\varepsilon {_n})t} \ME{\psi_{{\alpha_{\epsilon'}}}}{\hat{H}_{\rm int}(t)}{\psi_{\alpha_n}} \,   \mathcal  U_{(L_cl)LM}(\varepsilon {_n},t),
 \end{eqnarray}
is solved numerically in the \textit{amplitude coefficient equations method (ACE)}. Similar approach is applied, for example, in \cite{bello2021}. Here we extend the index $n$ to the continuum because   continuum states must be enumerated to solve the system (\ref{eq:diff}). $\hat{H}_{\rm int}(t)=A(t)\hat{D}$  is the product of the vector potential of the field and  {the} dipole operator in a velocity gauge.

Within the framework of {\sl nonstationary perturbation theory (PT)}, the expansion coefficients themselves are further expanded into series. Let  {u}s pick up some final energy $\varepsilon_f$ and consider three lowest orders of PT  after the end of pulse, so further we omit $t$ in the amplitudes.

For an unpolarized atom with an initial orbital angular momentum $L=0$, the first {-}order coefficients that describe direct ionization  to the mainlines (ML) by XUV components of the electric field:
\begin{align}\label{eq:ampl1}
& \mathcal  U^{(1)}_{(L_cl)LM}(\varepsilon_f)=\sum_\Lambda c_\Lambda\frac{1}{\sqrt{3}}\delta_{\Lambda M}D^{(1)}_{(L_cl)1}\,,\\
&D^{(1)}_{(L_cl)1}=-i\MEred{\varepsilon_{f};(L_cl)1}{\hat{D}}{\varepsilon_{0},\!0}\int_{-\tau/2}^{\tau/2} \hspace{-10pt} A_{\rm xuv}(t)e^{i(\varepsilon_f-\varepsilon_{0})t}dt\,.
\end{align}
Here $\MEred{}{}{}$ is a reduced dipole matrix element that does not depend on magnetic quantum numbers.
 In the first order, PT affects only states with $L=1$, and it is explicitly indicated in the equations.
Unlike our previous work \cite{popova2025}, here we introduce $c_\Lambda$ coefficient for the XUV components to allow more flexibility in choosing a pulse propagation direction. 

The second {-}order amplitudes describe absorption or emission of an IR photon leading to appearance of side\-bands (SB) through up- and down-energy transitions:

\begin{align}\label{eq:ampl2}
& \mathcal  U^{(2),\nu}_{(L_cl)LM}(\varepsilon_f)=\frac{(\pm1)^\lambda}{\sqrt{3}\hat{L}}\sum_{\lambda\Lambda}c_{\lambda}c_\Lambda\CGC{1}{\Lambda}{1}{\pm\lambda}{L}{M}D^{(2),\nu}_{(L_cl)L}\,,
\end{align}
\vspace{7pt}
\begin{align}
\label{eq:amplD2}
&D^{(2),\nu}_{(L_cl)L}=  
 \sum_{n}\!\MEred{\varepsilon_f,\!(L_cl)L}{D}{\varepsilon_{n},\!1}\MEred{\varepsilon_{n},\!1}{D}{\varepsilon_{0},\!0}\nonumber\\
&\hspace{7pt}\int_{-\tau}^{\tau}A_{\rm ir}^{u/d}(t)e^{i(\varepsilon_f-\varepsilon_{n})t} \int_{-\tau/2}^{t}\hspace{-10pt} A_{\rm xuv}(t')e^{i(\varepsilon_{n}-\varepsilon_{0})t'}dt'dt\,,
\end{align}

\noindent where  {the} `$+$' sign is for absorption amplitude ($\nu=u$, `up'), and  {the} `$-$' sign is for emission  ($\nu=d$, `down'). 

In equation~(\ref{eq:ampl2}), conventional notation for Clebsch\,---\,Gordan coefficients is used and $\hat{a}=\sqrt{2a+1}$. In the second order affects  {the} $n$th state with $L=0-2$.
 
Finally, the third-order amplitudes describe absorption or emission of two IR photons:
\begin{widetext}
\begin{align}
& \mathcal  U^{(3),\nu}_{(L_cl)LM}(\varepsilon_f)=\sum_{L_iM_i}\frac{(\pm1)^{\lambda}(\pm1)^{\lambda'}}{\sqrt{3}\hat{L}_i\hat{L}} 
\sum_{\lambda\lambda'\Lambda}c_{\lambda}c_{\lambda'}c_\Lambda\CGC{1}{\Lambda}{1}{\pm\lambda}{L_i}{M_i} \CGC{L_i}{M_i}{1}{\pm\lambda'}{L}{M}  D^{(3),\nu}_{L_i,(L_cl)L}\\
\label{eq:amplD3}
&D^{(3),\nu}_{L_i,(L_cl)L}=i   \sum_{k,n}\! \hspace{1pt}  \hat{l}_\gamma  \MEred{\varepsilon_f,(L_cl)L}{D}{\varepsilon_{n},L_i}\MEred{\varepsilon_{n},L_i}{D}{\varepsilon_{k},1} \MEred{\varepsilon_{k},1}{D}{\varepsilon_{0},\!0}\nonumber\\
&\hspace{110pt}\int_{-\tau}^{\tau}\!\!A_{\rm ir}^{u/d}(t)e^{i(\varepsilon-\varepsilon_{k})t} \int_{-\tau}^{t}\!\!\!A_{\rm ir}^{u/d}(t')e^{i(\varepsilon_{k}-\varepsilon_{n})t'} \int_{-\tau/2}^{t'} \hspace{-10pt}A_{\rm xuv}(t'')e^{i(\varepsilon_{k}-\varepsilon_{0})t''}dt''dt'dt \,.
\end{align}
\end{widetext}
There are four pathways of third-order amplitudes: $\nu=uu$ characterizes absorption of two IR photons, $\nu=dd$ ---  emission of two IR photons, while $\nu=ud$ and $\nu=du$ describe paths with one absorption and one emission of an IR photon. The last pathways return an electron to the energy of ML it starts of {f,} but their interference with first-order amplitudes does not invoke oscillations in the mainlines as these amplitudes do not depend on the IR phase (`$e^{-i(\omega t+\phi)}e^{i(\omega t+\phi)}$').

 {The ordered time integrals in Eqs. (\ref{eq:amplD2}), (\ref{eq:amplD3}) mean that the IR photon is absorbed after the XUV one, and the instantaneous intensity of the IR determines how long the electron remains in the MLs. For the later discussion, it is important that the instantaneous intensity of a linearly polarized field changes, but for a circularly polarized one it remains constant.}

The photo\-electron angular distribution (PAD) in PT and ACE is described as:
\begin{align}\label{eq:W}
&W(\varepsilon_f;\vartheta,\varphi)=\frac{1}{4\pi}\sum_{kq ll'LL'\atop nn'\nu\nu'}(-1)^{L_c+L+L'+k-M'}\hat{l}\hat{l}'\hat{L}\hat{L}'\nonumber\\
&\hspace{35pt}\CGC{l}{0}{l'}{0}{k}{0}\CGC{
L}{M}{L'}{-M'}{k}{q}\SechsJ{l}{L}{L_c}{L'}{l'}{k}\nonumber\\
&\hspace{10pt} \mathcal  U^{(n),\nu}_{(L_cl)LM}(\varepsilon_f) \mathcal  U^{(n'),\nu'\ast}_{(L_cl')L'M'}(\varepsilon_f) \frac{\sqrt{4\pi}}{\hat{k}}Y_{kq}(\theta,\varphi) \,,
\end{align}
where $n$ is an order of amplitude in PT. In the ACE, the resulting amplitude $\mathcal  U_{(L_cl)LM}(\varepsilon_f,t)$ is a converged infinite sum over $n$, and $\nu$ is a type of transition ($u,du,\dots$). In equation (\ref{eq:W}), conventional notations for Wigner $6j$-symbol and spherical harmonics are used \cite{polcor}. 

One can see that  equation (\ref{eq:W}) is formally the same as for the case of  {the} 1-SB scheme \cite{popova2025}. So, let us point out the key differences between  {the} 1-SB and 2-SB schemes:
\begin{enumerate}
\item In order to observe the IR field phase dependence of the photoelectron spectrum, one needs the interference of the pathways originating from different mainlines. The absorption of one IR photon from N-th ML brings an electron to the continuum at (N+1)th SB, while the emission of one IR photon from the subsequent (N+3)th ML brings an electron to the continuum at (N+2)th SB, therefore, the two-photon amplitudes from different mainlines end up at different energies and do not interfere. 
\item Angle-integrated spectra do not depend on the IR field phase{,} as all the IR phase-dependent interference terms are vanished due to parity conservation.
\item On the contrary, the interference of two- and three-order amplitudes (absorption of one IR photon from Nth ML and emission of two IR photons from (N+3)th ML bring {s} an electron to the continuum at (N+1)th SB), observable in the angle-resolved spectra, depends on the IR delay. 
\item Any allowed oscillations in the angle-resolved spectra occur at triple $3\omega_{\rm ir}$ frequency (`$e^{-i(\omega t+\phi)}(e^{i(\omega t+\phi)}e^{i(\omega t+\phi)})^*$') instead of the double $2\omega_{\rm ir}$ in conventional 1-SB RABBITT.
\item The  contributions to the angle-resolved spectra that  inherited symmetries from the electromagnetic field  ($k=0,2,4\dots$)  do not depend on IR field phase, while  contributions that violate the symmetry ($k=1,3,5\dots$) do depend. Unlike the case of  {the} 1-SB scheme, where the interfering terms are even and their ratios to angle-integrated photo\-ionization probability (term $k=0$) are not harmonic functions of IR phase, in  {the} 2-SB scheme, these ratios are harmonic functions. The latter makes extraction from the experimental data easier.
\end{enumerate}

\section{\label{sec:4} Computational aspects}

We use a neon atom as a target for the numerical calculations. The pulse (\ref{eq:field}) consists of the IR field with $\omega=1.55$~eV and peak field strength $E_{\rm{ir}}^0=2.5\cdot10^{-3}$ a.u. and its 15th, 18th{,} and 21st harmonics with equal peak field strength $E_{\rm{xuv}}^0=10^{-4}$ a.u{.} and $\tau=10$~fs.

To maintain consistency, the same spectroscopic model based on wave functions obtained within the MCHF package \cite{Fischer1997} was used as in \cite{popova2025}. Radial integrals $R_{l_il_f}(\varepsilon_{i}\varepsilon_{f})$ for transitions between continuum states were calculated using the divergence
elimination method \cite{Mercouris1994,Novikov2011}, under the assumption that any correlations between the free electron and the electrons of the residual ion can be neglected. These matrix elements consist of a regular ($\varepsilon_{i}\neq\varepsilon_{f}$) and a singular ($\varepsilon_{i}=\varepsilon_{f}$) terms, the latter behaves as $\sim\sqrt{2\varepsilon_{f}}$, so with the photoelectron energy increase it becomes more important.

To convert the radial integrals into the reduced matrix elements in the LS coupling scheme, a transition formula was used:
\begin{align}
\MEred{\varepsilon_{f},L_f}{D}{\varepsilon_{i},L_i}=(-1)^{Lc+L_i+1+l_i}\hat{l}_i\hat{L}_i\hat{L}_f\nonumber\\
\CGC{l_i}{0}{1}{0}{l_f}{0}\SechsJ{L_i}{l_i}{L_c}{l_f}{L_f}{1}R_{l_il_f}(\varepsilon_{i}\varepsilon_{f})\,.
\end{align}

The discrete–continuum and continuum–continuum dipole matrix elements vary slowly with energy, which allows them to be treated as constant over a sufficiently small  energy step  and to perform the continuum discretization procedure.

In the PT method, the time integrals were evaluated analytically, and the energy discretization step $d\varepsilon$ was set to 0.0011 a.u. Principal value integrals for continuum–continuum transitions were evaluated numerically over the energy interval   0.0011--0.4961 a.u. (0.03--13.5~eV).

In the ACE method, a double‑precision FORTRAN program was designed to integrate the system of ordinary differential equations for the expansion coefficients in the Coulomb basis. The 4th order Runge-Kutta method  was used to obtain initial values,  and the 3rd‑order Adams predictor-corrector method  \cite{MATHEMATICALHANDBOOK} was used to propagate the solution. {T}he integration was performed with a constant time step. The step size determines the relative error in the population of states (including states of the discretized continuum) and, accordingly, the norm of the total population. The step varied widely during the test calculations; its value, corresponding to approximately 50 points per period of the highest laser frequency, resulted in a relative norm error of about $10^{-4}$.

Most calculations were performed with $d\varepsilon=2.5\cdot10^{-3}$ a.u. 
Changing the energy step by a factor of 2 changed the computed quantities by no more than a few percent for pulse durations ($\pi\tau=15–60$ fs) and intensities  ($10^{11}–10^{13}$ W/cm$^2$), all typical for RABBITT experiments. Long pulses result in narrow photoelectron peaks and, therefore, require finer sampling step. Low IR intensity  ($<10^{11}$ W/cm$^2$) and short pulses (low flux) lead to deficient population in sidebands and larger statistical errors.  The considered range of continuum electron energies was $0.0025–0.625$ a.u. 
Note that decreasing the discretization step increases the computational cost of the problem $\sim N^2\approx \frac{1}{de^2}$, where N is the total number of energy points.

\section{\label{sec:5} results and discussion}

\widetext
\begin{figure*}[tb]
\includegraphics[width=0.8\textwidth]{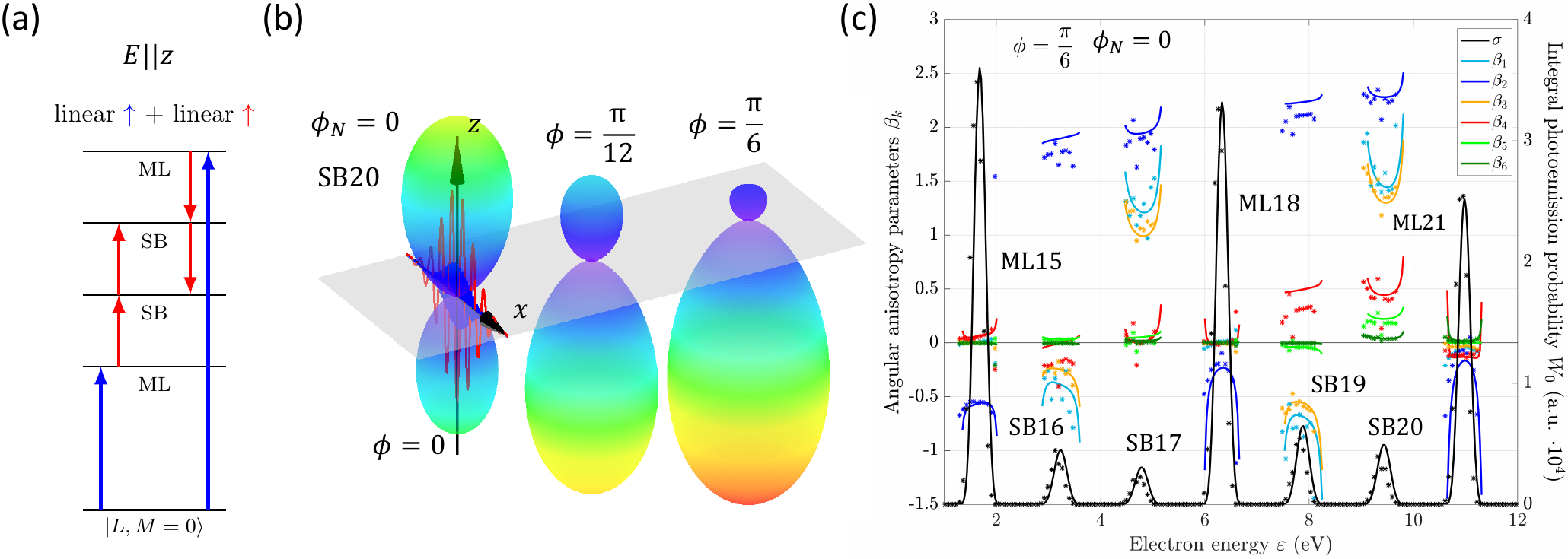} 
\caption{(a) The scheme of the 2-SB RABBITT for linearly ($E||z$) polarized fields; (b) PAD for different phases of the IR field $\phi$ in PT (note that it oscillates three times faster than the field); (c) the angular anisotropy parameters at $\phi={\pi}/{6}$ approximately corresponding to their maximum values for zero XUV phases $\phi_{N}=0$ and angle integrated photo\-electron spectra (does not depend on the phases).  }
\label{fig:linear1}
\end{figure*}
\endwidetext

For the case of linearly polarized in the same direction ($E||z$, Fig. \ref{fig:linear1}a) VUV comb and IR harmonic, the PAD is parametrized as follows:
\begin{eqnarray}
\label{eq:lin}
W^{ {\uparrow\uparrow}}(\theta;\varepsilon)&=& \frac{W_0^{ {\uparrow\uparrow}}}{4\pi}\left(1+\sum_{k=1,\dots,6} \beta^{ {\uparrow\uparrow}}_kP_{k}(\cos\theta)\right)\!,
\end{eqnarray}
where $P_k$  are the Legendre polynomials. Note that $k=6$ is the maximum value in 3 {rd}-order PT, in ACE, higher-order coefficients exist, but they are much smaller. The representation of Eq.~(\ref{eq:W}) in the form of (\ref{eq:lin}) is itself the definition of the integral photo\-emission probability $W_0$  ($k=0$) and the angular anisotropy parameters $\beta_{k}$. The PAD is axially symmetric with respect to the polarization direction. The phase-dependent ($\phi$) odd anisotropy parameters break the symmetry with respect to the plane  ($xy$)  orthogonal to the polarization.

 Simple physical explanation for symmetry breaking is as follows: the XUV field can be considered just as a pump which prepares axially symmetric ML states, and in a chosen SB a photo-/electron is promoted either by the interaction with one IR photon ($\cos(\omega t\pm\phi)$) or with two subsequent IR photons ($\cos^2(\omega t\mp\phi)$) [the plus sign is for emission, the minus sign is for absorption]. The total function of the IR field strength $\cos(\omega t)+\cos^2(\omega t)$ which determines the quantity of photo-/electrons is not symmetrical over period with respect to the $xy$ plane and depends on the phase, thus more photo-/electrons are emitted in one hemisphere than in the other. Also, as the instantaneous intensity of a linearly polarized field changes, the integral of this function depends on the phase, therefore, the phase allows for control of the form of the PAD.
 Analogous explanation was firstly given for bichromatic $\omega+2\omega$ scheme in \cite{Baranova1990}.

A typical PAD and its response to the  $\phi$ variation in this case are presented in Fig. \ref{fig:linear1}b for the photo\-electron energy corresponding to SB20 ($\varepsilon=20\omega-IP$, $IP$ is the ionization potential. The direction of the pulse oscillations is  shown schematically for $\phi=0$. When all $\phi_N$ are zero, the maximum asymmetry --- defined as the difference between the number of photoelectrons ejected into the upper and lower hemispheres --- is achieved near $\phi=\pi/6$. The corresponding angular anisotropy parameters are presented in Fig. \ref{fig:linear1}c.  In Fig. \ref{fig:linear1}c, points denote ACE results and lines denote PT results; thus a line connects points only when the two theories coincide exactly.  In the MLs, the overall probability and the angular anisotropy parameter $\beta_2$ behave very similarly to the well-known single-photon ionization of Ne:  {the} probability  decreases slowly{,} and $\beta_2$ starts from a negative value and increases. All other anisotropy parameters are minor because the single-photon process strongly prevails over the others. In the SBs, even anisotropy parameters are formed mainly by second-order contributions and  vary slowly along the considered energy region. It may seem surprising that the agreement between the theories is worse for SB16 and SB19 than for SB17 and SB20. The reason is the transitions from  a ML to a lower sideband are less intense than transitions to a higher sideband \cite{popova2023}, so SB16 and SB19  are described by PT less accurately than SB17 and SB20. The difference between the PT and ACE results (e.g., in $\beta_2$ for SB16 and SB19) is explained by the presence of multi\-photon processes of order higher than three  in ACE. The difference increases with the IR field strength increase and decreases with the decrease (within the limits of ACE stability, see above). 
The parameters $\beta_{1,3}$, caused by the interference of the amplitudes with different parities,   manifest a dramatic jump between successive sidebands. An electron emitted into the upper hemisphere at SB16 (SB19) and promoted further to SB17 (SB20) at the same time pumps the upper hemi\-sphere at SB17 (SB20) and depletes  {the} one at SB16 (SB19).

\begin{figure}
\includegraphics[width=0.49\textwidth]{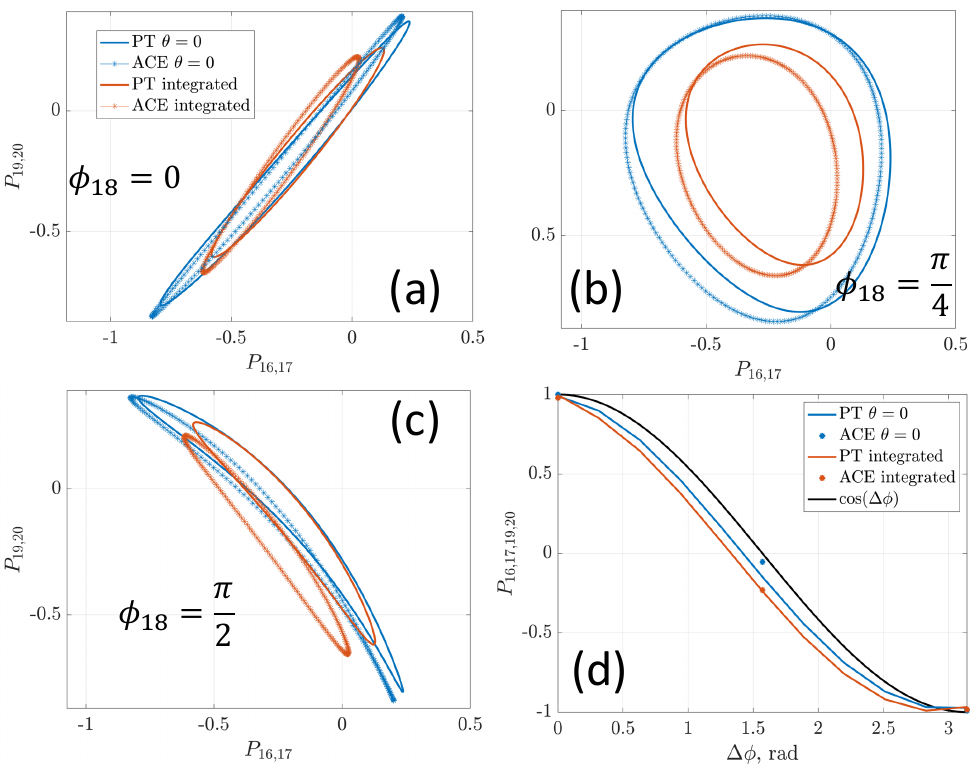}
\caption{Scheme `$\uparrow\uparrow$'. (a)--(c) Correlation plots between  $\mathcal P_{16,17}$ and $\mathcal P_{19,21}$ for three different phases of 18th harmonic; (d) The correlation function $\rho_{\rm 16, 17, 19, 20}$ in PT (solid lines) and ACE (dots) compared to a cosine function expected in the ideal conditions.  
 {It is necessary to note that, according to \cite{Maroju2020}, to characterize a pulse shape, one does not need to know individual phases of the harmonics. In 3rd order PT, odd $\beta$ parameters in the case of linearly polarized pulse depend on the phases of the subsequent harmonics in the following way: $\beta_k=|\beta_k|\cos(\arg{\beta_k}+\phi_{15}+\phi_{21}-2\phi_{18})$  {(see Appendix for details)}, therefore, $\phi_{18}=\pi/4$ is equivalent, for example, to $\phi_{15}=\pi/2$, $\phi_{15}=\pi/4$ and $\pi_{18}=\pi/8$.}}
\label{fig:linear2}
\end{figure}

For the considered target and energy range, the anisotropy parameters up to and including $\beta_4$ are significant enough to be extracted in experiments at  modern facilities (e.g., in \cite{moioli2025} both angle- and phase-differential anisotropy parameters were successfully obtained).

In Tab.~\ref{tab:placeholder}, the upper part lists contributions of different channels for linearly polarized fields and the lower part shows contributions for circularly polarized XUV and linearly polarized IR as they are two the most promising geometries. The even waves ($s\,,d^{\,1\!}P$) that appear in  {the} single-photon process dominate in MLs. Channels primarily formed in two- or three-photon process with decreasing of  {the} electron angular momentum $l$ ($p^{\,1\!}S\,,p^{\,1\!}D\,, d^{\,1\!}F$) are populated quite evenly throughout the sidebands/mainlines, whereas  channels formed with increasing of  {the} angular momentum tend to be populated in the absorption process ($f^{\,1\!}D\,,g^{\,1\!}F$). The most pronounced example is the relative population of $\varepsilon p$ (SB16$\approx$SB17) and $\varepsilon f$ (SB16$>$SB17) waves. Another demonstration is the three-photon $g^{\,1\!}F$ channel at SB17 and SB20, whose population is comparable to the two-photon $\varepsilon p$ channels. These tendencies were discussed in terms of Fano {'s} prosperity rule for continuum \cite{Busto2016}.

\begin{table}
    \centering
    \begin{tabular}{c|c|c|cc|c|cc|c}
     &  $\varepsilon l^{2S+1}L$  & ML15  & SB16 & SB17 & ML18 & SB19 & SB20 & ML21 \\
         \hline
1 & $\varepsilon s^1P$ & 9.8 & 0.26 & 0.33 & 7.4 & 0.54 & 0.71 &5.6 \\
1 & $\varepsilon d^1P$ &  16.2  \tikzmark{a} \tikzmark{b}&  0.53 & 1.29 & 16.6 \tikzmark{a2} \tikzmark{b2} & 0.89 & 1.81 & 14.7\\
3 &$\varepsilon d^1F$ & 0.40  & 0.17 & 0.14 & 0.37 & 0.28 & 0.27 & 0.55\\
3 &$\varepsilon g^1F$ & 0.44  & 0.27 &   \tikzmark{g} \tikzmark{h} 1.1 & 1.46 & 0.63 & \tikzmark{g2} \tikzmark{h2} 1.35 & 2.1\\
         2 &$\varepsilon p^1S$ &  & 2.8 & 3.2 &  & 4.0 & 3.6 & \\
         2 &$\varepsilon p^1D$ &  & 1.33 & 1.66 &  & 1.60 & 1.94 & \\
         2 &$\varepsilon f^1D$ &  & \tikzmark{c} \tikzmark{d}  \hspace{-5pt} 5.9 \tikzmark{e} \tikzmark{f}  & 3.9 &  & \tikzmark{c2} \tikzmark{d2} 6.7 \tikzmark{e2} \tikzmark{f2} &  5.2 & \\
         \hline
1 & $\varepsilon s^1P$ & 8.5 & 0.54 & 0.54 & 6.4 & 0.74 & 0.74 &5.1 \\
1 & $\varepsilon d^1P$ & 16.5 & 0.47 & 1.00 & 17.0 & 0.80 & 1.37 & 15.5 \\
3 & $\varepsilon d^1D$ & 1.26 & 0.43 & 0.64 & 1.37 & 0.61 & 0.84 & 1.84 \\
3 & $\varepsilon d^1F$ & 0.32 & 0.14 & 0.11 & 0.30 & 0.23 & 0.22 & 0.45 \\
3 & $\varepsilon g^1F$ & 0.36 & 0.22 & 0.89 & 1.18 & 0.51 & 1.1 & 1.72 \\
         2 & $\varepsilon p^1P$ &  & 4.0 & 4.3 &  & 4.6 & 4.4 & \\
         2 & $\varepsilon p^1D$ &  & 1.2 & 1.4 &  & 1.4 & 1.7 & \\
         2 & $\varepsilon f^1D$ &  & 5.1 & 3.3 &  & 5.8 & 4.5 & 
    \end{tabular}
\begin{tikzpicture}[overlay, remember picture, shorten >=.5pt, shorten <=.5pt, transform canvas={yshift=.25\baselineskip}]
    \draw [->, rounded corners,line width=0.75pt,red] ({pic cs:a}) -- ([xshift=.75pt]{pic cs:b}) -- ([xshift=-1.5pt]{pic cs:c}) -- ([xshift=.75pt]{pic cs:d});
    \draw [->, rounded corners,line width=0.75pt,red] ({pic cs:e}) -- ([xshift=.75pt]{pic cs:f}) -- ([xshift=-1.5pt]{pic cs:g}) -- ([xshift=.75pt]{pic cs:h});
    \draw [->, rounded corners,line width=0.75pt,red] ({pic cs:a2}) -- ([xshift=.75pt]{pic cs:b2}) -- ([xshift=-1.5pt]{pic cs:c2}) -- ([xshift=.75pt]{pic cs:d2});
    \draw [->, rounded corners,line width=0.75pt,red] ({pic cs:e2}) -- ([xshift=.75pt]{pic cs:f2}) -- ([xshift=-1.5pt]{pic cs:g2}) -- ([xshift=.75pt]{pic cs:h2});    
\end{tikzpicture}    
    \caption{The contribution of the channels ($10^{-3}$) for co-linearly polarized fields (upper part) and circularly polarized XUV comb and linearly polarized IR (lower part). The red arrows show two IR photon absorption paths.}
    \label{tab:placeholder}
\end{table}

The above results are for fixed phases $\phi_N=0$ of XUV harmonics.
When harmonics are generated on FEL or in other conditions where it is impossible to directly control the phase, a special trick called intensity correlation can be applied. As shown in \cite{Maroju2020}, sideband oscillations depend on both $\phi_N$ and $\phi$. The difference between two SBs, for example, SB16 and SB17, in a given direction  (for linearly polarized field it is convenient to set $\theta=0$) can be quantified as:
\begin{align}
\mathcal P_{\rm 16, 17}=\frac{W(0,\rm SB17)-W(0,\rm SB16)}{W(0,\rm SB17)+W(0,\rm SB16)}\,,\label{eq:Ptheta}
\end{align}
and for hemisphere-integrated case:
\begin{align}
\tilde{\mathcal P}_{\rm 16, 17}=\frac{\int\limits_0^{\frac{\pi}{2}}\left(W(\theta,\rm SB17)-W(\theta,\rm SB16)\right)\sin \theta d\theta}{\int\limits_0^{\frac{\pi}{2}}\left(W(\theta,\rm SB17)+W(\theta,\rm SB16)\right)\sin \theta d\theta}\,.
\label{eq:Pint}  
\end{align}

If phase-integrated intensities of sidebands are equal, correlation plots built from Eqs. (\ref{eq:Ptheta}) and (\ref{eq:Pint}) for four consecutive sideband pairs (two below a ML, and two above) form an ellipse which eccentricity depends on  $\phi_{N}$s and each experimental point determines the $\phi$ at which a shot was taken (see Supplementary Material of \cite{Maroju2020}). For the considered energy range near  the ionization threshold the situation is different as  continuum-continuum matrix elements between states with initial energy $\varepsilon'$ and final energy $\varepsilon$ are highly asymmetric  {\cite{popova2023}} relative to the line $\varepsilon=\varepsilon'$ [unlike for higher energy domain examined in \cite{Maroju2020}].
Consequently,   correlation plots between $\mathcal P_{\rm SB, SB'}$ and $\mathcal P_{\rm SB'', SB'''}$ do not form a perfect ellipse, but rather a quasi-ellipse (see Fig. \ref{fig:linear2}a). 
The angle-integrated correlations are less pronounced than angle-resolved ones; all curves are shifted toward the left-lower quadrant because the upper SBs are less intense.

One note should be made. To plot the correlation functions one needs to know angular anisotropy parameters at all values of the IR phase $\phi$. In order to reduce the number of calculations, we computed $\beta_k$s at three phases ($0, \pi/6, \pi/3$) and then fitted their periodic dependence as: 

\begin{align}
&\beta(\phi)=A\cos(3\phi+B),\\
&A=\frac{1}{2}\sqrt{\left(\beta(0)-2\beta(\frac{\pi}{6})+\beta(\frac{\pi}{3})\right)^2+\left(\beta(0)-\beta(\frac{\pi}{3})\right)^2}\,,\\
&B=\atantwo\left(\beta(0)-2\beta(\frac{\pi}{6})+\beta(\frac{\pi}{3}),\beta(0)-\beta(\frac{\pi}{3})\right)\,.
\end{align}

For four consecutive sidebands we define the correlation function:

\begin{align}\label{rhooo}
\rho_{\rm 16, 17, 19, 20}=\frac{\rm{cov}(\mathcal P_{\rm 16, 17}\mathcal P_{\rm 19, 20})}{\sigma(\mathcal P_{\rm 16, 17})\sigma(\mathcal P_{\rm 19, 20})}\,,
\end{align}

where $\rm{cov}$ is covariance and $\sigma$ is standard deviation. This function still practically follows  $\cos{\Delta\Phi}$, where $\Delta\Phi=\phi_{N-3}+\phi_{N+3}-2\phi_{N}$ (see Fig. \ref{fig:linear2}b  {and Appendix}). Therefore, equal phase-integrated sideband intensities  are not a crucial condition for determining relative phases of XUV harmonics from the correlations plots. The slight phase shift between a cosine function and $\rho_{\rm 16, 17, 19, 20}$ is attributed to nonsigular part of the radial integrals $R_{l_il_f}(\varepsilon_{i}\varepsilon_{f})$ between continuum states; artificial switching off this part (leaving only singular terms in calculation) eliminates the difference. As for higher photoelectron energies the singular terms are more important, one may conclude that the method is more applicable for higher energy ranges.

\begin{figure}[h!]
\includegraphics[width=0.32\textwidth]{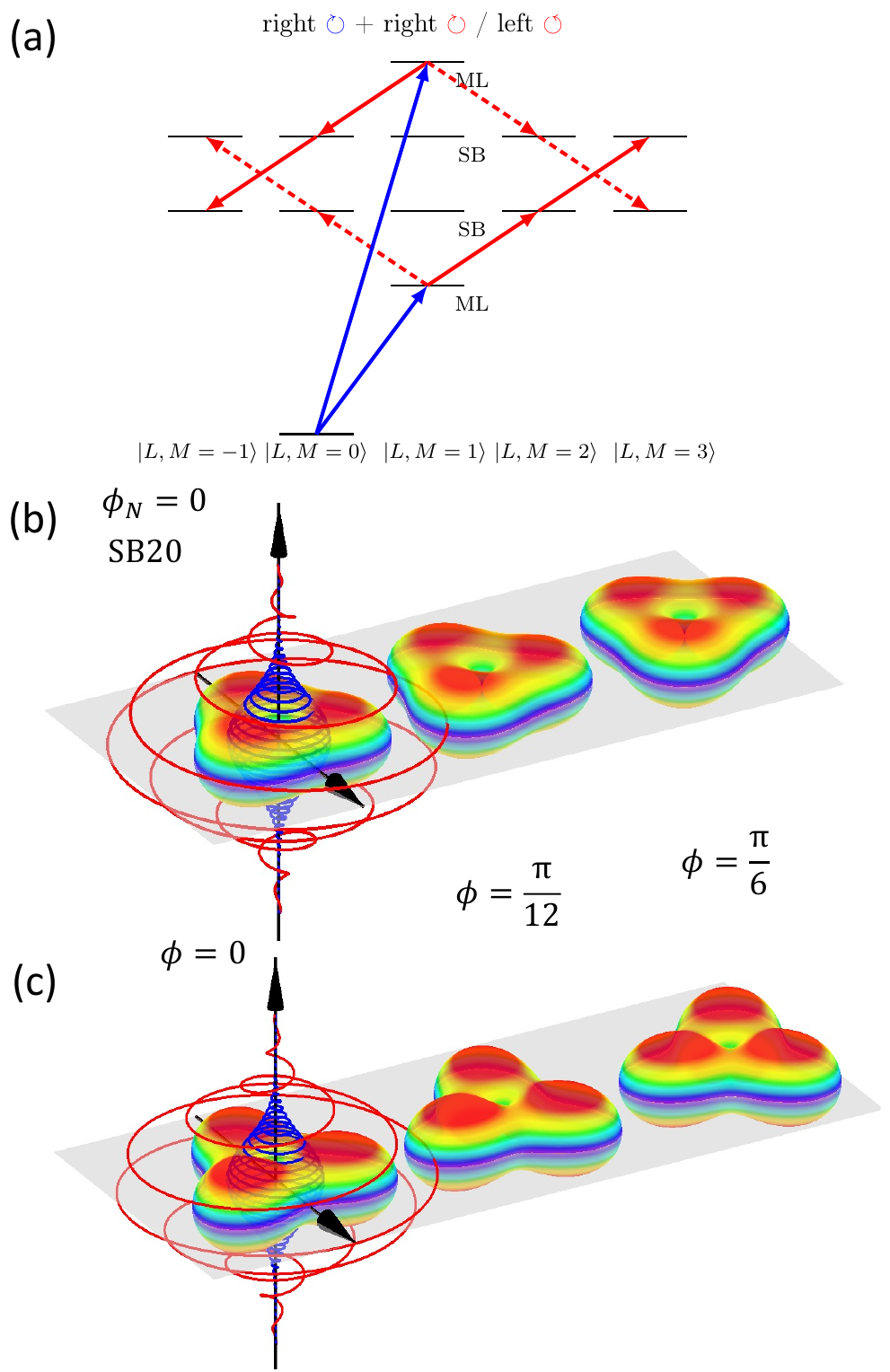} 
\caption{(a) Population of states of different magnetic quantum number, solid lines represent right circular polarization, dashed --- left circular polarization; PAD for different phases of the IR field $\phi$ in PT for right circularly polarized XUV pulse and right (b) and left (c) circularly polarized IR pulse. }
\label{fig:circular}
\end{figure}

The question  arises whether we can control the circularly polarized field using a similar technique. The selection rules governing the magnetic quantum number for circularly polarized harmonics exclude low $L$-channels, the corresponding scheme of allowed transitions is presented in Fig. \ref{fig:circular}a; $\Delta M$ for the absorption and emission branches is strictly 3. The PAD for the case of both right-polarized fields `$\circlearrowleft\circlearrowleft$' as well as for opposite helicities, `$\circlearrowleft\circlearrowright$', is parametrized as follows:

\begin{eqnarray}
\label{eq:rightright}
  &&W^{ {\circlearrowleft\circlearrowleft}}(\theta,\varphi;\varepsilon)=\frac{W_0^{ {\circlearrowleft\circlearrowleft}}}{4\pi}\Biggl(1+\sum_{k=2,4,6} \beta^{ {\circlearrowleft\circlearrowleft}}_kP_{k}(\cos\theta)+\nonumber\\
  &&\hspace{30pt}\sum_{3,5} |\beta^{ {\circlearrowleft\circlearrowleft}}_{k3}| P_k^3(\cos\theta)\cos(3\varphi-\text{arg}[\beta^{\circlearrowleft\circlearrowleft}_{k3}])\Biggl)\,.
\end{eqnarray}  

Here $P_k^q$ are Schmidt semi-normalized associated Legendre polynomials (for $q\ne0$ they differ by a factor $(-1)^q\sqrt{\frac{2(k-q)!}{(k+q)!}}$ from unnormalized functions). 
The choice of semi-normalized form of associated Legendre polynomials is made for the consistency of $\beta_{k,q=0}$ and $\beta_{k,q\ne0}$ values.  From Eq. (\ref{eq:rightright}) one can conclude that the PADs possess three-lobe $C$ symmetry, whose phase dependency  reduces to the rotation.

For neon,  {the} odd angular anisotropy parameter $\beta_{33}\leq0.25$ for the equal helicities  and  $\beta_{33}\leq0.5$ for opposite helicities. 
Three-lobe structure for both schemes is  clearly visible to the eye in theoretical depictions (presented in Fig.~\ref{fig:circular}b,c for SB20), but the evaluated asymmetry, defined as the difference between the number of electrons ejected along the lobe and in the opposite direction $[W(\pi/2,\phi)-W(\pi/2,\phi+\pi/3)]/[W(\pi/2,\phi)+W(\pi/2,\phi+\pi/3)]$, is small for `$\circlearrowleft\circlearrowleft$' case. However, the main experimental difficulty arises from  the PADs' three-lobe $D_{3h}$ symmetry itself: such patterns are inconvenient to detect  reliably with modern velocity map imaging (VMI) detectors.

The angular resolved circular magnetic dichroism is difficult to detect because it requires precise matching of the phase $\phi$ for $\circlearrowleft\circlearrowleft$ and  $\circlearrowleft\circlearrowright$ schemes. The angle-integrated circular magnetic dichroism for 2-SB scheme is modest (0.25) that slightly higher than for 1-SB scheme (0.17) but still difficult to be detected. As well as for 1-SB scheme,  special conditions like auto\-ionizing states needed to increase dichroism.   {It is important to remember that while absorption of a circularly polarized photon brings a screw to a system, the stimulated emission does the opposite, therefore, in 2-SB scheme, the difference of helicities for up- and down-pathways is 3. Such a large screw means that PAD is flattened to the $xy$-plane and emission of an electron along the field propagation direction is strongly suppressed.}

The same PAD parametrization is applicable to a case with circularly polarized IR field and linearly polarized in the direction of IR propagation XUV pulse (`$\uparrow\circlearrowleft$'). Unfortunately, for the chosen target, the odd anisotropy parameters are even smaller than for  {the} bi\-circular scheme, thus{,} we do not present this scheme. 

 {As XUV field only prepares a state, and for these three schemes ($\circlearrowleft\circlearrowleft$,  $\circlearrowleft\circlearrowright$ and $\uparrow\circlearrowleft$) the prepared states are axially symmetric and the IR field is the same, it is not surprising that PADs are parametrized equally. The three-lobe structure inherits the symmetry of the function of the IR field strength $(\cos(x\pm\phi)-i\sin(x\pm\phi))^2+(\cos(x\mp\phi)+i\sin(x\mp\phi))$; note that if one photon is absorbed, two are emitted or vice versa.  As the instantaneous intensity of a circularly  polarized field does not change, therefore, the phase control reduces to simple rotation  of the PAD.}

\widetext
\begin{figure*}
\includegraphics[width=0.79\textwidth]{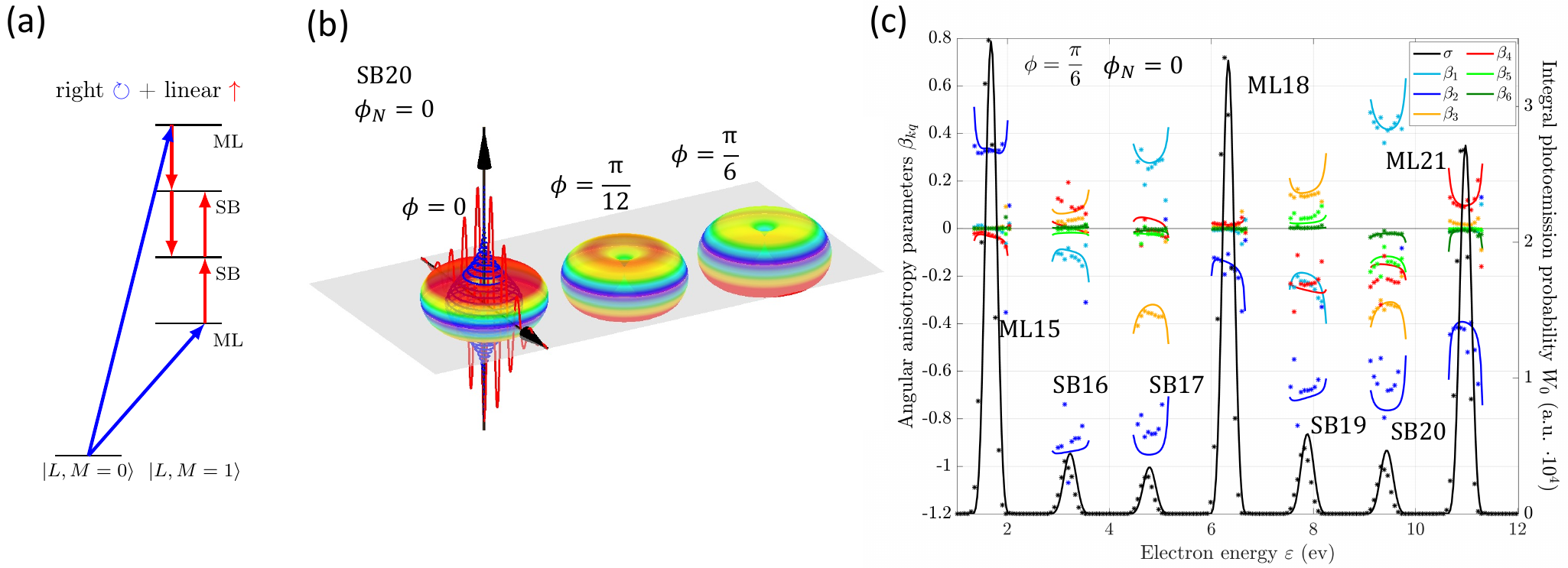} 
\caption{(a) The scheme of the 2-SB RABBITT for `$\circlearrowleft\uparrow$'-scheme (b) PAD for different phases of the IR field $\phi$ in PT (note that it oscillates three times faster than the field); (c) angular anisotropy parameters at $\phi={\pi}/{6}$ (near their maximum values for zero XUV phases $\phi_{N}=0$) and integrated photoelectron spectra (does not depend on phases).}
\label{fig:circlinear1}
\end{figure*}
\endwidetext

\begin{figure}[h!]
\includegraphics[width=0.49\textwidth]{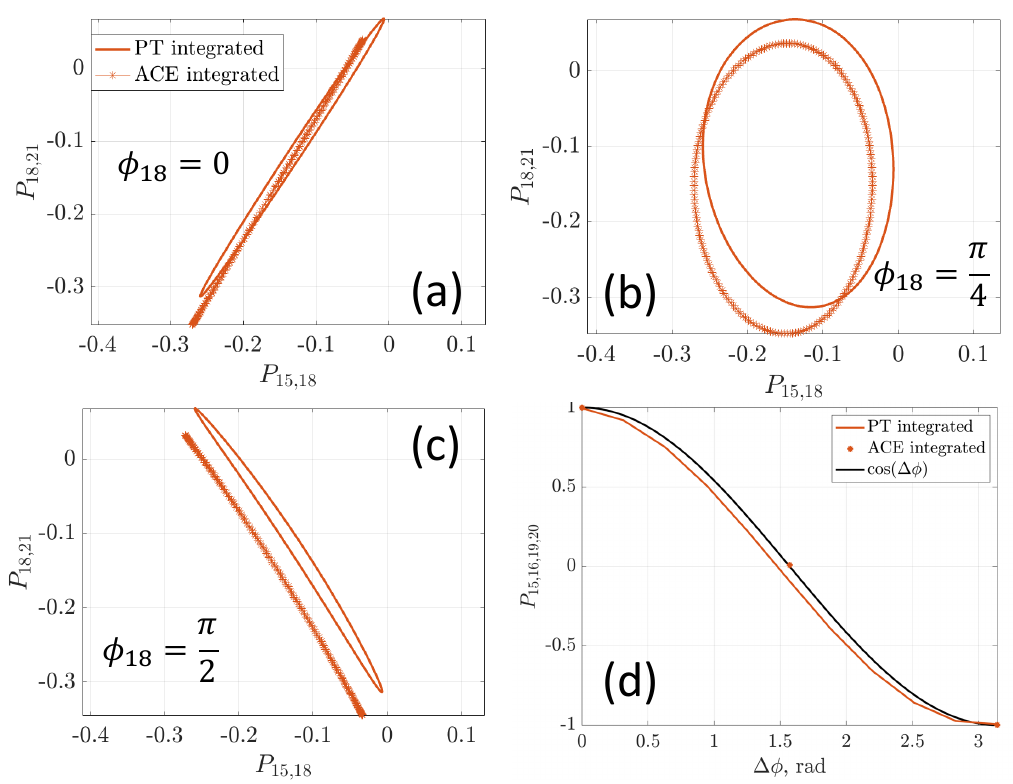}
\caption{Scheme `$\circlearrowleft\uparrow$'. (a)--(c) Correlation plots between  $\mathcal P_{16,17}$ and $\mathcal P_{19,21}$ for three different phases of 18th harmonic; (d)  The correlation function $\rho_{\rm 16, 17, 19, 20}$ in PT (solid lines) and ACE (dots) compared to a cosine function expected in the ideal conditions.  }
\label{fig:circlinear2}
\end{figure}

On the contrary, the scheme with  circularly polarized XUV field ($ \bm k||z$) and IR linearly polarized ($\bm E||z$), `$\circlearrowleft\uparrow$', is very promising. The PAD possesses axial symmetry with respect to the IR polarization vector:
\begin{eqnarray}    
  \label{eq:rightlin}
 W^{ {\circlearrowleft\uparrow}}(\theta;\varepsilon)&=& \frac{W_0^{ {\circlearrowleft\uparrow}}}{4\pi}\left(1+\sum_{k=1,\dots,6} \beta^{ {\circlearrowleft\uparrow}}_kP_{k}(\cos\theta)\right)\!.
\end{eqnarray}

A scheme showing the allowed transitions between states with different magnetic quantum number {s} is presented in Fig. \ref{fig:circlinear1}a, the corresponding PADs for the SB20 are shown in Fig. \ref{fig:circlinear1}b, and angular anisotropy parameters for the IR phase of their maximum values (applicable for the odd parameters) --- in Fig. \ref{fig:circlinear1}c.  {The symmetry for this geometry may look counterintuitive: one may expect that, as the circularly polarized field is involved, the phase control would reduce to a simple rotation. That was the case for a bi\-chromatic ionization \cite{Gryzlova19}. However, here, following the proposed interpretation, we can say that again the XUV field prepares axially symmetric ML states, and the linear IR field breaks the symmetry the way it does for $\uparrow\uparrow$ geometry.} The angular anisotropy parameters as a function of energy manifest  {a} tendency similar to the case of linearly polarized fields (Fig.~\ref{fig:linear1}c): `single photon' behavior of ML, minor variation of even parameters at SBs and a crucial jump of odd parameters between two adjacent SBs. The PAD demonstrates a donut-like pattern. The first absorption event brings a screw (orientation) to the system, and 
the subsequent absorption transfers this orientation to the SBs and, resulting in negative $\beta_2$ in the SBs.  Since negative $\beta_2$ suppresses emission of electrons along the quantization axis, that necessarily leads to either small or  opposite-signed odd $\beta$s   (see~Fig. \ref{fig:circlinear1}b,c). The simplest way to interpret the orientation transfer is to consider a hydrogen-like model: ionization from $1s$-shell promotes an electron to a $p$ state with $m=+1$ and subsequent ionization events conserve $m$; all wavefunctions $p+1$, $d+1${,} etc{.} are zero at $\theta=0$, therefore, $\beta_1+\beta_3+...=0$.

We can construct the correlation ellipses like the ones for linearly polarized fields (Eq.~\ref{eq:Pint}) for the angle-integrated spectrum, but not for $\theta=0,\pi$. For neon, they are expected to be about three times smaller (Fig. \ref{fig:circlinear2}a). The good news is that the correlation function is almost perfect (Fig. \ref{fig:circlinear2}b).

\begin{figure}[h!]
\includegraphics[width=0.35\textwidth]{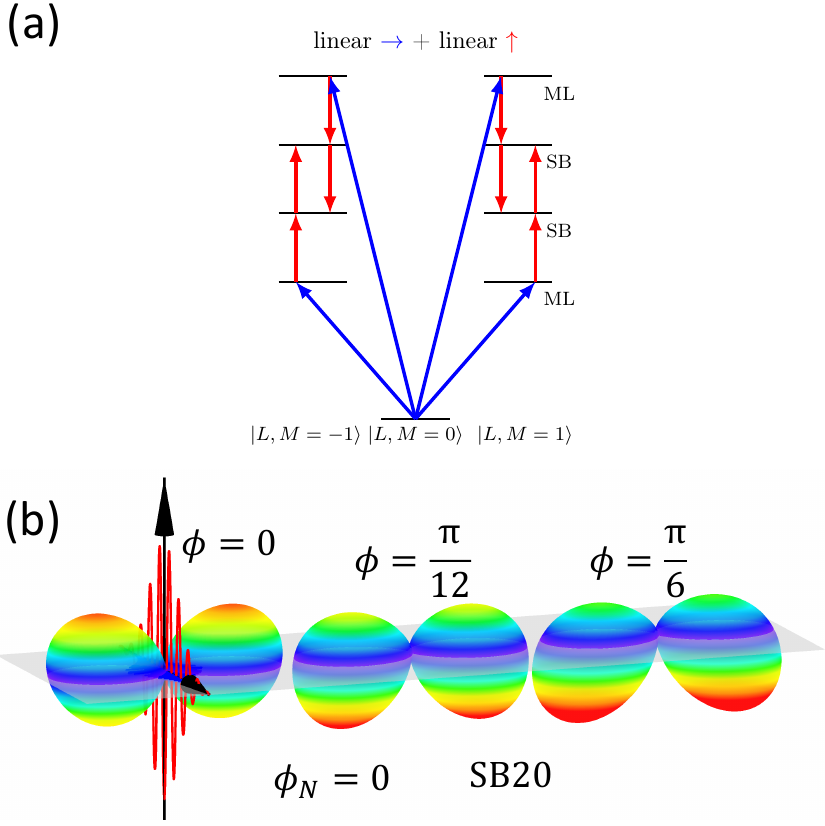} 
\caption{Linearly polarized fields in the perpendicular directions. (a) Population of states of different magnetic quantum number; (b) PAD for different phases of the IR field $\phi$ in PT. }
\label{fig:linearperp}
\end{figure}

Finally, we consider  {a} setup with crossed polarization vectors of XUV comb and IR harmonic, '$\rightarrow\uparrow$'. The scheme is one of the easiest to realize but possesses the lowest symmetry that makes it difficult to observe. The simplest parametrization with a minimal number of parameters is in the coordinate system $\bm E_{\rm ir}||z$ and $\bm E_{\rm xuv}||x$ (the corresponding  scheme is presented in Fig. \ref{fig:linearperp}a):

\begin{figure}[h!]
\includegraphics[width=0.4\textwidth]{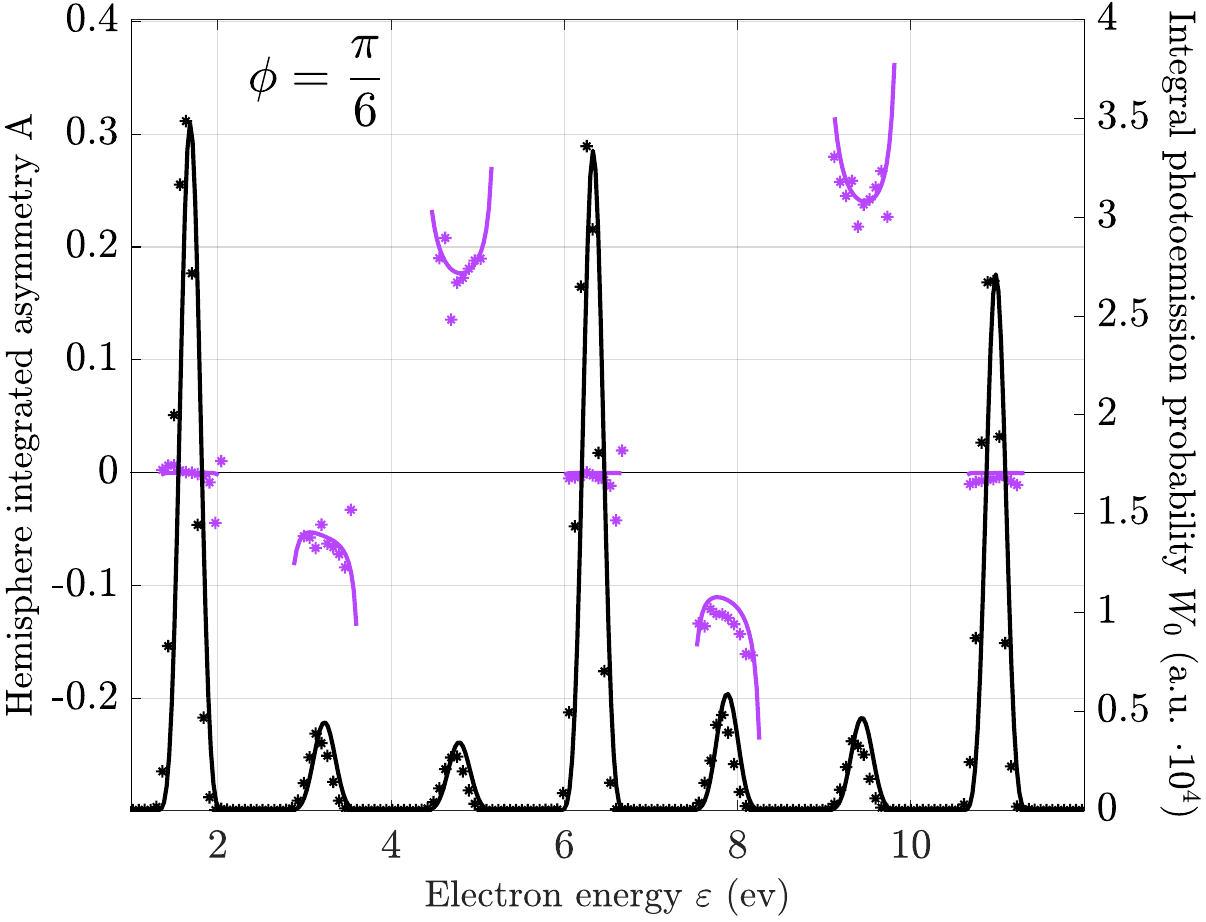} 
\caption{The asymmetry of electron emitted in the upper and bottom hemi\-sphere for crossed linearly polarized fields.}
\label{fig:linearperp_asym}
\end{figure}

\begin{align}
\label{eq:linperp}
   W^{ {\rightarrow\uparrow}}(\theta,\varphi;\varepsilon)=&\frac{W_0^{ {\rightarrow\uparrow}}}{4\pi}\Biggl(1+\sum_{k=1,\dots,6} \beta^{ {\rightarrow\uparrow}}_kP_{k}(\cos\theta)+\nonumber\\
   &\cos(2\varphi)\sum_{k=2,\dots,6}\beta^{ {\rightarrow\uparrow}}_{k2}P_k^2(\cos\theta)\Biggl)\,. 
\end{align}

There is no axial symmetry, but there are two symmetry planes: $xz$ and $yz$. The odd parameters break  the symmetry with respect to the $xy$ plane. The corresponding PADs are presented in Fig. \ref{fig:linearperp}b.  { Here the XUV field prepares the states with axis of symmetry orthogonal to quantization axis forming “8” in the $xy$ plane, and the linear IR field breaks the up/down symmetry the way it does for $\uparrow\uparrow$ geometry.} In spite of lower symmetry, the PAD noticeably varies with the phase of the IR field. Because of the large number of anisotropy parameters{,} from a practical point of view{,} it is constructive to consider a more robust quantity such as hemi\-sphere-integrated asymmetry Fig.~\ref{fig:linearperp_asym}.

Only $\beta_k0$ parameters contribute to the hemi\-sphere-integrated asymmetry. Its value is measurable, so in some experimental conditions one could prefer this scheme to '$\uparrow\uparrow$'.

\section{Conclusion}

In the paper, we  investigated  {the} 2-SB RABBITT scheme, concentrating on  effects connected to the pulse polarization. 
 {From the experimental point of view, the key difference of this scheme from traditional 1-SB RABBITT is the fact that angle-integrated cross-sections do not depend on the IR field phase. It may be helpful in accounting for intensity fluctuations during the experiment as one can make a correction keeping that in mind. Due to the same reason, the odd angular anisotropy parameters become harmonic functions of the IR field phase, which also helps in their extraction.}

We analyzed six configurations of highly symmetrical electric field strength  patterns. The considered polarization geometries are: collinear and orthogonal linearly polarized components, circularly polarized components of the same and opposite helicities, and combination of linearly and circularly polarized components. We found that three of them, i.e.{,}  circular polarization of equal helicities, crossed linear and circular XUV with linear seed produce PADs that are completely different from those in bi\-chromatic $\omega+2\omega$ ionization, whose most important feature is also parity mixing. 

We found out that two of them lead to axially symmetrical PADs: besides the well-know {n} scheme with both the IR seed and the XUV comb linearly polarized in the same direction (${\uparrow\uparrow}$),  a circularly polarized  XUV comb propagating along the IR field polarization vector $E_{\rm ir}$ (${\circlearrowleft\uparrow}$)  also does not depend on the azimuthal angle $\phi$ (quantization axis $z||E_{\rm ir}$). Therefore, the ${\circlearrowleft\uparrow}$ scheme may be useful for experimental characterization of the circularly polarized XUV fields.

For the schemes with axially symmetrical PADs (${\uparrow\uparrow}$ and ${\circlearrowleft\uparrow}$), we constructed correlation functions between two pairs of subsequent sidebands and showed that the XUV harmonics' phases can be reconstructed regardless of whether the condition of equal sidebands {'} intensities is fulfilled.

For setups with circularly polarized IR field, independently on polarization of XUV component (${\circlearrowleft\circlearrowleft}$ and ${\uparrow\circlearrowleft}$), PAD pos {s}esses a three-lobe structure, and its variation with the IR phase reduces to a rotation. There is no significant circular magnetic dichroism between the ${\circlearrowleft\circlearrowleft}$ and ${\circlearrowleft\circlearrowright}$ schemes in a flat-continuum region despite a major difference in allowed channels.

Finally, for crossed linearly polarized IR and XUV components (${\leftarrow\uparrow}$), PAD has only two symmetry planes, however the planes differ from those in bi\-chromatic ionization. For this scheme, variation of the IR phase changes the PAD significantly{,} keeping the symmetry planes steady.

 {Further perspective of the work may lie in studying the properties of the RABBITT scheme in the vicinity of autoionization state, particularly studying  circular magnetic dichroism to explore dipole-forbidden states.}

\section*{ACKNOWLEDGMENTS}

The part of the study connected to the realization of the amplitude coefficient equations (ACE) method was conducted under the state assignment of Lomonosov Moscow State University. 
The part of the research related to the theoretical description and numerical simulations in the perturbation theory (PT) approach was supported by the Russian Science Foundation (Project No. 25–72–10172).

\section*{Appendix}
Let us mark fourth  sequential SBs as 'a', 'b', 'c' and 'd'. A signal is fitted as: 
\begin{eqnarray}
W^{i}&=&S_0^{i}+S_2^{i}\cos(3\omega t+\phi^{i})\,, i=a\,,b\,,c\,,d.
\end{eqnarray}
Parameters $S_0\,,S_2\,,\phi$ depend not only energy, but also if the signal angle-integrated or at a particular angle. Then
\begin{eqnarray}
\mathcal{P}_{ab}&=&S^a_0-S_0^b+S_2^{ab}\cos(3\omega+\psi_{ab})\,;\\
S_2^{ab}&=&\sqrt{|S_2^a|^2+|S_2^b|^2-2S_2^aS_0^b\cos(\phi^a-\phi^b)}\,,\\
\cos\psi_{ab}&=&\frac{S_2^a\cos\phi^a-S_2^b\cos\phi^b}{\sqrt{|S_2^a|^2+|S_2^b|^2-2S_2^aS_0^b\cos(\phi^a-\phi^b)}}\,.
\end{eqnarray}
If $S_2^a\approx S_2^b$, which is a natural assumption because of population conservation, then
\begin{eqnarray}
\cos\psi_{ab}&=&\frac{\cos\phi^a-\cos\phi^b}{\sqrt{2-2\cos(\phi^a-\phi^b)}}=-\frac{\sin\frac{\phi^a+\phi^b}{2}}{\sin\frac{\phi^a-\phi^b}{2}}\,,
\end{eqnarray}
\begin{eqnarray}
\tan\psi_{ab}&=&-\frac{\sin\phi^a-\sin\phi^b}{\cos\phi^a-\cos\phi^b}=-\frac{\cos\frac{\phi^a+\phi^b}{2}}{\sin\frac{\phi^a+\phi^b}{2}}\,,\\
\psi_{ab}&=&\frac{\pi+\phi^a+\phi^b}{2}\,.
\end{eqnarray}
\begin{eqnarray}
\rho_{abcd}&=&\frac{\int \cos(3\omega+\psi_{ab})\cos(3\omega+\psi_{cd})}{\sqrt{\int\cos^2(3\omega+\psi_{ab})\int\cos^2(3\omega+\psi_{cd})}}=\\
&=&\cos(\psi_{ab}-\psi_{cd})=\cos(\frac{\phi^a+\phi^b-\phi^c-\phi^d}{2})
\end{eqnarray}
The argument of the cosine function in the last equation, neglecting the tiny atomic phase differences (see discussion below Eq. \ref{rhooo}), is equal $\phi_{N}+\phi_{N+6}-2\phi_{N+3}$.

\bibliography{references}

\begin{thebibliography}{58}%
\makeatletter
\providecommand \@ifxundefined [1]{%
 \@ifx{#1\undefined}
}%
\providecommand \@ifnum [1]{%
 \ifnum #1\expandafter \@firstoftwo
 \else \expandafter \@secondoftwo
 \fi
}%
\providecommand \@ifx [1]{%
 \ifx #1\expandafter \@firstoftwo
 \else \expandafter \@secondoftwo
 \fi
}%
\providecommand \natexlab [1]{#1}%
\providecommand \enquote  [1]{``#1''}%
\providecommand \bibnamefont  [1]{#1}%
\providecommand \bibfnamefont [1]{#1}%
\providecommand \citenamefont [1]{#1}%
\providecommand \href@noop [0]{\@secondoftwo}%
\providecommand \href [0]{\begingroup \@sanitize@url \@href}%
\providecommand \@href[1]{\@@startlink{#1}\@@href}%
\providecommand \@@href[1]{\endgroup#1\@@endlink}%
\providecommand \@sanitize@url [0]{\catcode `\\12\catcode `\$12\catcode
  `\&12\catcode `\#12\catcode `\^12\catcode `\_12\catcode `\%12\relax}%
\providecommand \@@startlink[1]{}%
\providecommand \@@endlink[0]{}%
\providecommand \url  [0]{\begingroup\@sanitize@url \@url }%
\providecommand \@url [1]{\endgroup\@href {#1}{\urlprefix }}%
\providecommand \urlprefix  [0]{URL }%
\providecommand \Eprint [0]{\href }%
\providecommand \doibase [0]{https://doi.org/}%
\providecommand \selectlanguage [0]{\@gobble}%
\providecommand \bibinfo  [0]{\@secondoftwo}%
\providecommand \bibfield  [0]{\@secondoftwo}%
\providecommand \translation [1]{[#1]}%
\providecommand \BibitemOpen [0]{}%
\providecommand \bibitemStop [0]{}%
\providecommand \bibitemNoStop [0]{.\EOS\space}%
\providecommand \EOS [0]{\spacefactor3000\relax}%
\providecommand \BibitemShut  [1]{\csname bibitem#1\endcsname}%
\let\auto@bib@innerbib\@empty
\bibitem [{\citenamefont {Maroju}\ \emph {et~al.}(2020)\citenamefont {Maroju},
  \citenamefont {Grazioli}, \citenamefont {Di~Fraia}, \citenamefont {Moioli},
  \citenamefont {Ertel}, \citenamefont {Ahmadi}, \citenamefont {Plekan},
  \citenamefont {Finetti}, \citenamefont {Allaria}, \citenamefont {Giannessi},
  \citenamefont {De~Ninno}, \citenamefont {Spezzani}, \citenamefont {Penco},
  \citenamefont {Spampinati}, \citenamefont {Demidovich}, \citenamefont
  {Danailov}, \citenamefont {Borghes}, \citenamefont {Kourousias},
  \citenamefont {Sanches Dos~Reis}, \citenamefont {Bill{\'e}}, \citenamefont
  {Lutman}, \citenamefont {Squibb}, \citenamefont {Feifel}, \citenamefont
  {Carpeggiani}, \citenamefont {Reduzzi}, \citenamefont {Mazza}, \citenamefont
  {Meyer}, \citenamefont {Bengtsson}, \citenamefont {Ibrakovic}, \citenamefont
  {Simpson}, \citenamefont {Mauritsson}, \citenamefont {Csizmadia},
  \citenamefont {Dumergue}, \citenamefont {K{\"u}hn}, \citenamefont
  {Nandiga~Gopalakrishna}, \citenamefont {You}, \citenamefont {Ueda},
  \citenamefont {Labeye}, \citenamefont {B{\ae}kh{\o}j}, \citenamefont
  {Schafer}, \citenamefont {Gryzlova}, \citenamefont {Grum-Grzhimailo},
  \citenamefont {Prince}, \citenamefont {Callegari},\ and\ \citenamefont
  {Sansone}}]{Maroju2020}%
  \BibitemOpen
  \bibfield  {author} {\bibinfo {author} {\bibfnamefont {P.~K.}\ \bibnamefont
  {Maroju}}, \bibinfo {author} {\bibfnamefont {C.}~\bibnamefont {Grazioli}},
  \bibinfo {author} {\bibfnamefont {M.}~\bibnamefont {Di~Fraia}}, \bibinfo
  {author} {\bibfnamefont {M.}~\bibnamefont {Moioli}}, \bibinfo {author}
  {\bibfnamefont {D.}~\bibnamefont {Ertel}}, \bibinfo {author} {\bibfnamefont
  {H.}~\bibnamefont {Ahmadi}}, \bibinfo {author} {\bibfnamefont
  {O.}~\bibnamefont {Plekan}}, \bibinfo {author} {\bibfnamefont
  {P.}~\bibnamefont {Finetti}}, \bibinfo {author} {\bibfnamefont
  {E.}~\bibnamefont {Allaria}}, \bibinfo {author} {\bibfnamefont
  {L.}~\bibnamefont {Giannessi}}, \bibinfo {author} {\bibfnamefont
  {G.}~\bibnamefont {De~Ninno}}, \bibinfo {author} {\bibfnamefont
  {C.}~\bibnamefont {Spezzani}}, \bibinfo {author} {\bibfnamefont
  {G.}~\bibnamefont {Penco}}, \bibinfo {author} {\bibfnamefont
  {S.}~\bibnamefont {Spampinati}}, \bibinfo {author} {\bibfnamefont
  {A.}~\bibnamefont {Demidovich}}, \bibinfo {author} {\bibfnamefont {M.~B.}\
  \bibnamefont {Danailov}}, \bibinfo {author} {\bibfnamefont {R.}~\bibnamefont
  {Borghes}}, \bibinfo {author} {\bibfnamefont {G.}~\bibnamefont {Kourousias}},
  \bibinfo {author} {\bibfnamefont {C.~E.}\ \bibnamefont {Sanches Dos~Reis}},
  \bibinfo {author} {\bibfnamefont {F.}~\bibnamefont {Bill{\'e}}}, \bibinfo
  {author} {\bibfnamefont {A.~A.}\ \bibnamefont {Lutman}}, \bibinfo {author}
  {\bibfnamefont {R.~J.}\ \bibnamefont {Squibb}}, \bibinfo {author}
  {\bibfnamefont {R.}~\bibnamefont {Feifel}}, \bibinfo {author} {\bibfnamefont
  {P.}~\bibnamefont {Carpeggiani}}, \bibinfo {author} {\bibfnamefont
  {M.}~\bibnamefont {Reduzzi}}, \bibinfo {author} {\bibfnamefont
  {T.}~\bibnamefont {Mazza}}, \bibinfo {author} {\bibfnamefont
  {M.}~\bibnamefont {Meyer}}, \bibinfo {author} {\bibfnamefont
  {S.}~\bibnamefont {Bengtsson}}, \bibinfo {author} {\bibfnamefont
  {N.}~\bibnamefont {Ibrakovic}}, \bibinfo {author} {\bibfnamefont {E.~R.}\
  \bibnamefont {Simpson}}, \bibinfo {author} {\bibfnamefont {J.}~\bibnamefont
  {Mauritsson}}, \bibinfo {author} {\bibfnamefont {T.}~\bibnamefont
  {Csizmadia}}, \bibinfo {author} {\bibfnamefont {M.}~\bibnamefont {Dumergue}},
  \bibinfo {author} {\bibfnamefont {S.}~\bibnamefont {K{\"u}hn}}, \bibinfo
  {author} {\bibfnamefont {H.}~\bibnamefont {Nandiga~Gopalakrishna}}, \bibinfo
  {author} {\bibfnamefont {D.}~\bibnamefont {You}}, \bibinfo {author}
  {\bibfnamefont {K.}~\bibnamefont {Ueda}}, \bibinfo {author} {\bibfnamefont
  {M.}~\bibnamefont {Labeye}}, \bibinfo {author} {\bibfnamefont {J.~E.}\
  \bibnamefont {B{\ae}kh{\o}j}}, \bibinfo {author} {\bibfnamefont {K.~J.}\
  \bibnamefont {Schafer}}, \bibinfo {author} {\bibfnamefont {E.~V.}\
  \bibnamefont {Gryzlova}}, \bibinfo {author} {\bibfnamefont {A.~N.}\
  \bibnamefont {Grum-Grzhimailo}}, \bibinfo {author} {\bibfnamefont {K.~C.}\
  \bibnamefont {Prince}}, \bibinfo {author} {\bibfnamefont {C.}~\bibnamefont
  {Callegari}},\ and\ \bibinfo {author} {\bibfnamefont {G.}~\bibnamefont
  {Sansone}},\ }\bibfield  {title} {\bibinfo {title} {Attosecond pulse shaping
  using a seeded free-electron laser},\ }\href
  {https://doi.org/10.1038/s41586-020-2005-6} {\bibfield  {journal} {\bibinfo
  {journal} {Nature}\ }\textbf {\bibinfo {volume} {578}},\ \bibinfo {pages}
  {386} (\bibinfo {year} {2020})}\BibitemShut {NoStop}%
\bibitem [{\citenamefont {Ritchie}(1976)}]{Burke}%
  \BibitemOpen
  \bibfield  {author} {\bibinfo {author} {\bibfnamefont {B.}~\bibnamefont
  {Ritchie}},\ }\bibfield  {title} {\bibinfo {title} {Theory of the angular
  distribution of photoelectrons ejected from optically active molecules and
  molecular negative ions},\ }\href {https://doi.org/10.1103/PhysRevA.13.1411}
  {\bibfield  {journal} {\bibinfo  {journal} {Phys. Rev. A}\ }\textbf {\bibinfo
  {volume} {13}},\ \bibinfo {pages} {1411} (\bibinfo {year}
  {1976})}\BibitemShut {NoStop}%
\bibitem [{\citenamefont {Fano}(1969)}]{Fano}%
  \BibitemOpen
  \bibfield  {author} {\bibinfo {author} {\bibfnamefont {U.}~\bibnamefont
  {Fano}},\ }\bibfield  {title} {\bibinfo {title} {Spin orientation of
  photoelectrons ejected by circularly polarized light},\ }\href
  {https://doi.org/10.1103/PhysRev.178.131} {\bibfield  {journal} {\bibinfo
  {journal} {Phys. Rev.}\ }\textbf {\bibinfo {volume} {178}},\ \bibinfo {pages}
  {131} (\bibinfo {year} {1969})}\BibitemShut {NoStop}%
\bibitem [{\citenamefont {Lewenstein}\ \emph {et~al.}(1994)\citenamefont
  {Lewenstein}, \citenamefont {Balcou}, \citenamefont {Ivanov}, \citenamefont
  {L'Huillier},\ and\ \citenamefont {Corkum}}]{Lewenstein1994}%
  \BibitemOpen
  \bibfield  {author} {\bibinfo {author} {\bibfnamefont {M.}~\bibnamefont
  {Lewenstein}}, \bibinfo {author} {\bibfnamefont {P.}~\bibnamefont {Balcou}},
  \bibinfo {author} {\bibfnamefont {M.~Y.}\ \bibnamefont {Ivanov}}, \bibinfo
  {author} {\bibfnamefont {A.}~\bibnamefont {L'Huillier}},\ and\ \bibinfo
  {author} {\bibfnamefont {P.~B.}\ \bibnamefont {Corkum}},\ }\bibfield  {title}
  {\bibinfo {title} {Theory of high-harmonic generation by low-frequency laser
  fields},\ }\href {https://doi.org/10.1103/PhysRevA.49.2117} {\bibfield
  {journal} {\bibinfo  {journal} {Phys. Rev. A}\ }\textbf {\bibinfo {volume}
  {49}},\ \bibinfo {pages} {2117} (\bibinfo {year} {1994})}\BibitemShut
  {NoStop}%
\bibitem [{\citenamefont {Strelkov}\ \emph {et~al.}(2016)\citenamefont
  {Strelkov}, \citenamefont {Platonenko}, \citenamefont {Sterzhantov},\ and\
  \citenamefont {Ryabikin}}]{strelkov2016}%
  \BibitemOpen
  \bibfield  {author} {\bibinfo {author} {\bibfnamefont {V.~V.}\ \bibnamefont
  {Strelkov}}, \bibinfo {author} {\bibfnamefont {V.~T.}\ \bibnamefont
  {Platonenko}}, \bibinfo {author} {\bibfnamefont {A.~F.}\ \bibnamefont
  {Sterzhantov}},\ and\ \bibinfo {author} {\bibfnamefont {M.~Y.}\ \bibnamefont
  {Ryabikin}},\ }\bibfield  {title} {\bibinfo {title} {Attosecond
  electromagnetic pulses: generation, measurement, and application. generation
  of high-order harmonics of an intense laser field for attosecond pulse
  production},\ }\href {https://doi.org/10.3367/UFNe.2015.12.037670} {\bibfield
   {journal} {\bibinfo  {journal} {Physics-Uspekhi}\ }\textbf {\bibinfo
  {volume} {59}},\ \bibinfo {pages} {425} (\bibinfo {year} {2016})}\BibitemShut
  {NoStop}%
\bibitem [{\citenamefont {Callegari}\ \emph {et~al.}(2021)\citenamefont
  {Callegari}, \citenamefont {Grum-Grzhimailo}, \citenamefont {Ishikawa},
  \citenamefont {Prince}, \citenamefont {Sansone},\ and\ \citenamefont
  {Ueda}}]{CALLEGARI20211}%
  \BibitemOpen
  \bibfield  {author} {\bibinfo {author} {\bibfnamefont {C.}~\bibnamefont
  {Callegari}}, \bibinfo {author} {\bibfnamefont {A.~N.}\ \bibnamefont
  {Grum-Grzhimailo}}, \bibinfo {author} {\bibfnamefont {K.~L.}\ \bibnamefont
  {Ishikawa}}, \bibinfo {author} {\bibfnamefont {K.~C.}\ \bibnamefont
  {Prince}}, \bibinfo {author} {\bibfnamefont {G.}~\bibnamefont {Sansone}},\
  and\ \bibinfo {author} {\bibfnamefont {K.}~\bibnamefont {Ueda}},\ }\bibfield
  {title} {\bibinfo {title} {Atomic, molecular and optical physics applications
  of longitudinally coherent and narrow bandwidth free-electron lasers},\
  }\href {https://doi.org/https://doi.org/10.1016/j.physrep.2020.12.002}
  {\bibfield  {journal} {\bibinfo  {journal} {Physics Reports}\ }\textbf
  {\bibinfo {volume} {904}},\ \bibinfo {pages} {1} (\bibinfo {year}
  {2021})}\BibitemShut {NoStop}%
\bibitem [{\citenamefont {Allaria}\ \emph {et~al.}(2014)\citenamefont
  {Allaria}, \citenamefont {Diviacco}, \citenamefont {Callegari}, \citenamefont
  {Finetti}, \citenamefont {Mahieu}, \citenamefont {Viefhaus}, \citenamefont
  {Zangrando}, \citenamefont {De~Ninno}, \citenamefont {Lambert}, \citenamefont
  {Ferrari}, \citenamefont {Buck}, \citenamefont {Ilchen}, \citenamefont
  {Vodungbo}, \citenamefont {Mahne}, \citenamefont {Svetina}, \citenamefont
  {Spezzani}, \citenamefont {Di~Mitri}, \citenamefont {Penco}, \citenamefont
  {Trov\'o}, \citenamefont {Fawley}, \citenamefont {Rebernik}, \citenamefont
  {Gauthier}, \citenamefont {Grazioli}, \citenamefont {Coreno}, \citenamefont
  {Ressel}, \citenamefont {Kivim\"aki}, \citenamefont {Mazza}, \citenamefont
  {Glaser}, \citenamefont {Scholz}, \citenamefont {Seltmann}, \citenamefont
  {Gessler}, \citenamefont {Gr\"unert}, \citenamefont {De~Fanis}, \citenamefont
  {Meyer}, \citenamefont {Knie}, \citenamefont {Moeller}, \citenamefont
  {Raimondi}, \citenamefont {Capotondi}, \citenamefont {Pedersoli},
  \citenamefont {Plekan}, \citenamefont {Danailov}, \citenamefont {Demidovich},
  \citenamefont {Nikolov}, \citenamefont {Abrami}, \citenamefont {Gautier},
  \citenamefont {L\"uning}, \citenamefont {Zeitoun},\ and\ \citenamefont
  {Giannessi}}]{Allaria2014}%
  \BibitemOpen
  \bibfield  {author} {\bibinfo {author} {\bibfnamefont {E.}~\bibnamefont
  {Allaria}}, \bibinfo {author} {\bibfnamefont {B.}~\bibnamefont {Diviacco}},
  \bibinfo {author} {\bibfnamefont {C.}~\bibnamefont {Callegari}}, \bibinfo
  {author} {\bibfnamefont {P.}~\bibnamefont {Finetti}}, \bibinfo {author}
  {\bibfnamefont {B.}~\bibnamefont {Mahieu}}, \bibinfo {author} {\bibfnamefont
  {J.}~\bibnamefont {Viefhaus}}, \bibinfo {author} {\bibfnamefont
  {M.}~\bibnamefont {Zangrando}}, \bibinfo {author} {\bibfnamefont
  {G.}~\bibnamefont {De~Ninno}}, \bibinfo {author} {\bibfnamefont
  {G.}~\bibnamefont {Lambert}}, \bibinfo {author} {\bibfnamefont
  {E.}~\bibnamefont {Ferrari}}, \bibinfo {author} {\bibfnamefont
  {J.}~\bibnamefont {Buck}}, \bibinfo {author} {\bibfnamefont {M.}~\bibnamefont
  {Ilchen}}, \bibinfo {author} {\bibfnamefont {B.}~\bibnamefont {Vodungbo}},
  \bibinfo {author} {\bibfnamefont {N.}~\bibnamefont {Mahne}}, \bibinfo
  {author} {\bibfnamefont {C.}~\bibnamefont {Svetina}}, \bibinfo {author}
  {\bibfnamefont {C.}~\bibnamefont {Spezzani}}, \bibinfo {author}
  {\bibfnamefont {S.}~\bibnamefont {Di~Mitri}}, \bibinfo {author}
  {\bibfnamefont {G.}~\bibnamefont {Penco}}, \bibinfo {author} {\bibfnamefont
  {M.}~\bibnamefont {Trov\'o}}, \bibinfo {author} {\bibfnamefont {W.~M.}\
  \bibnamefont {Fawley}}, \bibinfo {author} {\bibfnamefont {P.~R.}\
  \bibnamefont {Rebernik}}, \bibinfo {author} {\bibfnamefont {D.}~\bibnamefont
  {Gauthier}}, \bibinfo {author} {\bibfnamefont {C.}~\bibnamefont {Grazioli}},
  \bibinfo {author} {\bibfnamefont {M.}~\bibnamefont {Coreno}}, \bibinfo
  {author} {\bibfnamefont {B.}~\bibnamefont {Ressel}}, \bibinfo {author}
  {\bibfnamefont {A.}~\bibnamefont {Kivim\"aki}}, \bibinfo {author}
  {\bibfnamefont {T.}~\bibnamefont {Mazza}}, \bibinfo {author} {\bibfnamefont
  {L.}~\bibnamefont {Glaser}}, \bibinfo {author} {\bibfnamefont
  {F.}~\bibnamefont {Scholz}}, \bibinfo {author} {\bibfnamefont
  {J.}~\bibnamefont {Seltmann}}, \bibinfo {author} {\bibfnamefont
  {P.}~\bibnamefont {Gessler}}, \bibinfo {author} {\bibfnamefont
  {J.}~\bibnamefont {Gr\"unert}}, \bibinfo {author} {\bibfnamefont
  {A.}~\bibnamefont {De~Fanis}}, \bibinfo {author} {\bibfnamefont
  {M.}~\bibnamefont {Meyer}}, \bibinfo {author} {\bibfnamefont
  {A.}~\bibnamefont {Knie}}, \bibinfo {author} {\bibfnamefont {S.~P.}\
  \bibnamefont {Moeller}}, \bibinfo {author} {\bibfnamefont {L.}~\bibnamefont
  {Raimondi}}, \bibinfo {author} {\bibfnamefont {F.}~\bibnamefont {Capotondi}},
  \bibinfo {author} {\bibfnamefont {E.}~\bibnamefont {Pedersoli}}, \bibinfo
  {author} {\bibfnamefont {O.}~\bibnamefont {Plekan}}, \bibinfo {author}
  {\bibfnamefont {M.~B.}\ \bibnamefont {Danailov}}, \bibinfo {author}
  {\bibfnamefont {A.}~\bibnamefont {Demidovich}}, \bibinfo {author}
  {\bibfnamefont {I.}~\bibnamefont {Nikolov}}, \bibinfo {author} {\bibfnamefont
  {A.}~\bibnamefont {Abrami}}, \bibinfo {author} {\bibfnamefont
  {J.}~\bibnamefont {Gautier}}, \bibinfo {author} {\bibfnamefont
  {J.}~\bibnamefont {L\"uning}}, \bibinfo {author} {\bibfnamefont
  {P.}~\bibnamefont {Zeitoun}},\ and\ \bibinfo {author} {\bibfnamefont
  {L.}~\bibnamefont {Giannessi}},\ }\bibfield  {title} {\bibinfo {title}
  {Control of the polarization of a vacuum-ultraviolet, high-gain,
  free-electron laser},\ }\href {https://doi.org/10.1103/PhysRevX.4.041040}
  {\bibfield  {journal} {\bibinfo  {journal} {Phys. Rev. X}\ }\textbf {\bibinfo
  {volume} {4}},\ \bibinfo {pages} {041040} (\bibinfo {year}
  {2014})}\BibitemShut {NoStop}%
\bibitem [{\citenamefont {von Korff~Schmising}\ \emph
  {et~al.}(2017)\citenamefont {von Korff~Schmising}, \citenamefont {Weder},
  \citenamefont {Noll}, \citenamefont {Pfau}, \citenamefont {Hennecke},
  \citenamefont {Strüber}, \citenamefont {Radu}, \citenamefont {Schneider},
  \citenamefont {Staeck}, \citenamefont {Günther}, \citenamefont {Lüning},
  \citenamefont {Merhe}, \citenamefont {Buck}, \citenamefont {Hartmann},
  \citenamefont {Viefhaus}, \citenamefont {Treusch},\ and\ \citenamefont
  {Eisebitt}}]{Schmising2017}%
  \BibitemOpen
  \bibfield  {author} {\bibinfo {author} {\bibfnamefont {C.}~\bibnamefont {von
  Korff~Schmising}}, \bibinfo {author} {\bibfnamefont {D.}~\bibnamefont
  {Weder}}, \bibinfo {author} {\bibfnamefont {T.}~\bibnamefont {Noll}},
  \bibinfo {author} {\bibfnamefont {B.}~\bibnamefont {Pfau}}, \bibinfo {author}
  {\bibfnamefont {M.}~\bibnamefont {Hennecke}}, \bibinfo {author}
  {\bibfnamefont {C.}~\bibnamefont {Strüber}}, \bibinfo {author}
  {\bibfnamefont {I.}~\bibnamefont {Radu}}, \bibinfo {author} {\bibfnamefont
  {M.}~\bibnamefont {Schneider}}, \bibinfo {author} {\bibfnamefont
  {S.}~\bibnamefont {Staeck}}, \bibinfo {author} {\bibfnamefont {C.~M.}\
  \bibnamefont {Günther}}, \bibinfo {author} {\bibfnamefont {J.}~\bibnamefont
  {Lüning}}, \bibinfo {author} {\bibfnamefont {A.~e.~d.}\ \bibnamefont
  {Merhe}}, \bibinfo {author} {\bibfnamefont {J.}~\bibnamefont {Buck}},
  \bibinfo {author} {\bibfnamefont {G.}~\bibnamefont {Hartmann}}, \bibinfo
  {author} {\bibfnamefont {J.}~\bibnamefont {Viefhaus}}, \bibinfo {author}
  {\bibfnamefont {R.}~\bibnamefont {Treusch}},\ and\ \bibinfo {author}
  {\bibfnamefont {S.}~\bibnamefont {Eisebitt}},\ }\bibfield  {title} {\bibinfo
  {title} {Generating circularly polarized radiation in the extreme ultraviolet
  spectral range at the free-electron laser flash},\ }\href
  {https://doi.org/10.1063/1.4983056} {\bibfield  {journal} {\bibinfo
  {journal} {Review of Scientific Instruments}\ }\textbf {\bibinfo {volume}
  {88}},\ \bibinfo {pages} {053903} (\bibinfo {year} {2017})}\BibitemShut
  {NoStop}%
\bibitem [{\citenamefont {Kfir}\ \emph {et~al.}(2015)\citenamefont {Kfir},
  \citenamefont {Grychtol}, \citenamefont {Turgut}, \citenamefont {Knut},
  \citenamefont {Zusin}, \citenamefont {Popmintchev}, \citenamefont
  {Popmintchev}, \citenamefont {Nembach}, \citenamefont {Shaw}, \citenamefont
  {Fleischer}, \citenamefont {Kapteyn}, \citenamefont {Murnane},\ and\
  \citenamefont {Cohen}}]{Kfir2015}%
  \BibitemOpen
  \bibfield  {author} {\bibinfo {author} {\bibfnamefont {O.}~\bibnamefont
  {Kfir}}, \bibinfo {author} {\bibfnamefont {P.}~\bibnamefont {Grychtol}},
  \bibinfo {author} {\bibfnamefont {E.}~\bibnamefont {Turgut}}, \bibinfo
  {author} {\bibfnamefont {R.}~\bibnamefont {Knut}}, \bibinfo {author}
  {\bibfnamefont {D.}~\bibnamefont {Zusin}}, \bibinfo {author} {\bibfnamefont
  {D.}~\bibnamefont {Popmintchev}}, \bibinfo {author} {\bibfnamefont
  {T.}~\bibnamefont {Popmintchev}}, \bibinfo {author} {\bibfnamefont
  {H.}~\bibnamefont {Nembach}}, \bibinfo {author} {\bibfnamefont {J.~M.}\
  \bibnamefont {Shaw}}, \bibinfo {author} {\bibfnamefont {A.}~\bibnamefont
  {Fleischer}}, \bibinfo {author} {\bibfnamefont {H.}~\bibnamefont {Kapteyn}},
  \bibinfo {author} {\bibfnamefont {M.}~\bibnamefont {Murnane}},\ and\ \bibinfo
  {author} {\bibfnamefont {O.}~\bibnamefont {Cohen}},\ }\bibfield  {title}
  {\bibinfo {title} {Generation of bright phase-matched circularly-polarized
  extreme ultraviolet high harmonics},\ }\href
  {https://doi.org/10.1038/nphoton.2014.293} {\bibfield  {journal} {\bibinfo
  {journal} {Nature Photonics}\ }\textbf {\bibinfo {volume} {9}},\ \bibinfo
  {pages} {99} (\bibinfo {year} {2015})}\BibitemShut {NoStop}%
\bibitem [{\citenamefont {Mahieu}\ \emph {et~al.}(2018)\citenamefont {Mahieu},
  \citenamefont {Stremoukhov}, \citenamefont {Gauthier}, \citenamefont
  {Spezzani}, \citenamefont {Alves}, \citenamefont {Vodungbo}, \citenamefont
  {Zeitoun}, \citenamefont {Malka}, \citenamefont {De~Ninno},\ and\
  \citenamefont {Lambert}}]{Mahieu2018}%
  \BibitemOpen
  \bibfield  {author} {\bibinfo {author} {\bibfnamefont {B.}~\bibnamefont
  {Mahieu}}, \bibinfo {author} {\bibfnamefont {S.}~\bibnamefont {Stremoukhov}},
  \bibinfo {author} {\bibfnamefont {D.}~\bibnamefont {Gauthier}}, \bibinfo
  {author} {\bibfnamefont {C.}~\bibnamefont {Spezzani}}, \bibinfo {author}
  {\bibfnamefont {C.}~\bibnamefont {Alves}}, \bibinfo {author} {\bibfnamefont
  {B.}~\bibnamefont {Vodungbo}}, \bibinfo {author} {\bibfnamefont
  {P.}~\bibnamefont {Zeitoun}}, \bibinfo {author} {\bibfnamefont
  {V.}~\bibnamefont {Malka}}, \bibinfo {author} {\bibfnamefont
  {G.}~\bibnamefont {De~Ninno}},\ and\ \bibinfo {author} {\bibfnamefont
  {G.}~\bibnamefont {Lambert}},\ }\bibfield  {title} {\bibinfo {title} {Control
  of ellipticity in high-order harmonic generation driven by two linearly
  polarized fields},\ }\href {https://doi.org/10.1103/PhysRevA.97.043857}
  {\bibfield  {journal} {\bibinfo  {journal} {Phys. Rev. A}\ }\textbf {\bibinfo
  {volume} {97}},\ \bibinfo {pages} {043857} (\bibinfo {year}
  {2018})}\BibitemShut {NoStop}%
\bibitem [{\citenamefont {Khokhlova}\ \emph {et~al.}(2021)\citenamefont
  {Khokhlova}, \citenamefont {Emelin}, \citenamefont {Ryabikin},\ and\
  \citenamefont {Strelkov}}]{Khokhlova2021}%
  \BibitemOpen
  \bibfield  {author} {\bibinfo {author} {\bibfnamefont {M.~A.}\ \bibnamefont
  {Khokhlova}}, \bibinfo {author} {\bibfnamefont {M.~Y.}\ \bibnamefont
  {Emelin}}, \bibinfo {author} {\bibfnamefont {M.~Y.}\ \bibnamefont
  {Ryabikin}},\ and\ \bibinfo {author} {\bibfnamefont {V.~V.}\ \bibnamefont
  {Strelkov}},\ }\bibfield  {title} {\bibinfo {title} {Polarization control of
  quasimonochromatic xuv light produced via resonant high-order harmonic
  generation},\ }\href {https://doi.org/10.1103/PhysRevA.103.043114} {\bibfield
   {journal} {\bibinfo  {journal} {Phys. Rev. A}\ }\textbf {\bibinfo {volume}
  {103}},\ \bibinfo {pages} {043114} (\bibinfo {year} {2021})}\BibitemShut
  {NoStop}%
\bibitem [{\citenamefont {Emelin}\ and\ \citenamefont
  {Ryabikin}(2025)}]{Emelin}%
  \BibitemOpen
  \bibfield  {author} {\bibinfo {author} {\bibfnamefont {M.}~\bibnamefont
  {Emelin}}\ and\ \bibinfo {author} {\bibfnamefont {M.}~\bibnamefont
  {Ryabikin}},\ }\bibfield  {title} {\bibinfo {title} {High-ellipticity
  resonant below-threshold harmonic generation by a helium atom driven by a
  moderately intense elliptically polarized laser field.},\ }\href
  {https://doi.org/doi.org/10.1007/s11082-025-08355-1} {\bibfield  {journal}
  {\bibinfo  {journal} {Opt Quant Electron}\ }\textbf {\bibinfo {volume}
  {57}},\ \bibinfo {pages} {434} (\bibinfo {year} {2025})}\BibitemShut
  {NoStop}%
\bibitem [{\citenamefont {Grum-Grzhimailo}\ \emph {et~al.}(2015)\citenamefont
  {Grum-Grzhimailo}, \citenamefont {Gryzlova}, \citenamefont {Staroselskaya},
  \citenamefont {Venzke},\ and\ \citenamefont {Bartschat}}]{grum2015}%
  \BibitemOpen
  \bibfield  {author} {\bibinfo {author} {\bibfnamefont {A.~N.}\ \bibnamefont
  {Grum-Grzhimailo}}, \bibinfo {author} {\bibfnamefont {E.~V.}\ \bibnamefont
  {Gryzlova}}, \bibinfo {author} {\bibfnamefont {E.~I.}\ \bibnamefont
  {Staroselskaya}}, \bibinfo {author} {\bibfnamefont {J.}~\bibnamefont
  {Venzke}},\ and\ \bibinfo {author} {\bibfnamefont {K.}~\bibnamefont
  {Bartschat}},\ }\bibfield  {title} {\bibinfo {title} {Interfering one-photon
  and two-photon ionization by femtosecond vuv pulses in the region of an
  intermediate resonance},\ }\href {https://doi.org/10.1103/PhysRevA.91.063418}
  {\bibfield  {journal} {\bibinfo  {journal} {Phys. Rev. A}\ }\textbf {\bibinfo
  {volume} {91}},\ \bibinfo {pages} {063418} (\bibinfo {year}
  {2015})}\BibitemShut {NoStop}%
\bibitem [{\citenamefont {Prince}\ \emph {et~al.}(2016)\citenamefont {Prince},
  \citenamefont {Allaria}, \citenamefont {Callegari}, \citenamefont {Cucini},
  \citenamefont {Ninno}, \citenamefont {Mitri},\ and\ \citenamefont
  {ithers}}]{NATPHOT-Prince-2016}%
  \BibitemOpen
  \bibfield  {author} {\bibinfo {author} {\bibfnamefont {K.~C.}\ \bibnamefont
  {Prince}}, \bibinfo {author} {\bibfnamefont {E.}~\bibnamefont {Allaria}},
  \bibinfo {author} {\bibfnamefont {C.}~\bibnamefont {Callegari}}, \bibinfo
  {author} {\bibfnamefont {R.}~\bibnamefont {Cucini}}, \bibinfo {author}
  {\bibfnamefont {G.~D.}\ \bibnamefont {Ninno}}, \bibinfo {author}
  {\bibfnamefont {S.~D.}\ \bibnamefont {Mitri}},\ and\ \bibinfo {author}
  {\bibnamefont {ithers}},\ }\bibfield  {title} {\bibinfo {title} {Coherent
  control with a short-wavelength free-electron laser},\ }\href@noop {}
  {\bibfield  {journal} {\bibinfo  {journal} {Nat. Photonics}\ }\textbf
  {\bibinfo {volume} {10}},\ \bibinfo {pages} {176} (\bibinfo {year}
  {2016})}\BibitemShut {NoStop}%
\bibitem [{\citenamefont {Frolov}\ \emph {et~al.}(2010)\citenamefont {Frolov},
  \citenamefont {Manakov}, \citenamefont {Silaev},\ and\ \citenamefont
  {Vvedenskii}}]{Frolov2010}%
  \BibitemOpen
  \bibfield  {author} {\bibinfo {author} {\bibfnamefont {M.~V.}\ \bibnamefont
  {Frolov}}, \bibinfo {author} {\bibfnamefont {N.~L.}\ \bibnamefont {Manakov}},
  \bibinfo {author} {\bibfnamefont {A.~A.}\ \bibnamefont {Silaev}},\ and\
  \bibinfo {author} {\bibfnamefont {N.~V.}\ \bibnamefont {Vvedenskii}},\
  }\bibfield  {title} {\bibinfo {title} {Analytic description of high-order
  harmonic generation by atoms in a two-color laser field},\ }\href
  {https://doi.org/10.1103/PhysRevA.81.063407} {\bibfield  {journal} {\bibinfo
  {journal} {Phys. Rev. A}\ }\textbf {\bibinfo {volume} {81}},\ \bibinfo
  {pages} {063407} (\bibinfo {year} {2010})}\BibitemShut {NoStop}%
\bibitem [{\citenamefont {Mancuso}\ \emph {et~al.}(2016)\citenamefont
  {Mancuso}, \citenamefont {Hickstein}, \citenamefont {Dorney}, \citenamefont
  {Ellis}, \citenamefont {Hasovi\ifmmode~\acute{c}\else \'{c}\fi{}},
  \citenamefont {Knut}, \citenamefont {Grychtol}, \citenamefont {Gentry},
  \citenamefont {Gopalakrishnan}, \citenamefont {Zusin}, \citenamefont
  {Dollar}, \citenamefont {Tong}, \citenamefont {Milo\ifmmode \check{s}\else
  \v{s}\fi{}evi\ifmmode~\acute{c}\else \'{c}\fi{}}, \citenamefont {Becker},
  \citenamefont {Kapteyn},\ and\ \citenamefont {Murnane}}]{Mancuso2016}%
  \BibitemOpen
  \bibfield  {author} {\bibinfo {author} {\bibfnamefont {C.~A.}\ \bibnamefont
  {Mancuso}}, \bibinfo {author} {\bibfnamefont {D.~D.}\ \bibnamefont
  {Hickstein}}, \bibinfo {author} {\bibfnamefont {K.~M.}\ \bibnamefont
  {Dorney}}, \bibinfo {author} {\bibfnamefont {J.~L.}\ \bibnamefont {Ellis}},
  \bibinfo {author} {\bibfnamefont {E.}~\bibnamefont
  {Hasovi\ifmmode~\acute{c}\else \'{c}\fi{}}}, \bibinfo {author} {\bibfnamefont
  {R.}~\bibnamefont {Knut}}, \bibinfo {author} {\bibfnamefont {P.}~\bibnamefont
  {Grychtol}}, \bibinfo {author} {\bibfnamefont {C.}~\bibnamefont {Gentry}},
  \bibinfo {author} {\bibfnamefont {M.}~\bibnamefont {Gopalakrishnan}},
  \bibinfo {author} {\bibfnamefont {D.}~\bibnamefont {Zusin}}, \bibinfo
  {author} {\bibfnamefont {F.~J.}\ \bibnamefont {Dollar}}, \bibinfo {author}
  {\bibfnamefont {X.-M.}\ \bibnamefont {Tong}}, \bibinfo {author}
  {\bibfnamefont {D.~B.}\ \bibnamefont {Milo\ifmmode \check{s}\else
  \v{s}\fi{}evi\ifmmode~\acute{c}\else \'{c}\fi{}}}, \bibinfo {author}
  {\bibfnamefont {W.}~\bibnamefont {Becker}}, \bibinfo {author} {\bibfnamefont
  {H.~C.}\ \bibnamefont {Kapteyn}},\ and\ \bibinfo {author} {\bibfnamefont
  {M.~M.}\ \bibnamefont {Murnane}},\ }\bibfield  {title} {\bibinfo {title}
  {Controlling electron-ion rescattering in two-color circularly polarized
  femtosecond laser fields},\ }\href
  {https://doi.org/10.1103/PhysRevA.93.053406} {\bibfield  {journal} {\bibinfo
  {journal} {Phys. Rev. A}\ }\textbf {\bibinfo {volume} {93}},\ \bibinfo
  {pages} {053406} (\bibinfo {year} {2016})}\BibitemShut {NoStop}%
\bibitem [{\citenamefont {Douguet}\ \emph {et~al.}(2016)\citenamefont
  {Douguet}, \citenamefont {Grum-Grzhimailo}, \citenamefont {Gryzlova},
  \citenamefont {Staroselskaya}, \citenamefont {Venzke},\ and\ \citenamefont
  {Bartschat}}]{douguet2016}%
  \BibitemOpen
  \bibfield  {author} {\bibinfo {author} {\bibfnamefont {N.}~\bibnamefont
  {Douguet}}, \bibinfo {author} {\bibfnamefont {A.~N.}\ \bibnamefont
  {Grum-Grzhimailo}}, \bibinfo {author} {\bibfnamefont {E.~V.}\ \bibnamefont
  {Gryzlova}}, \bibinfo {author} {\bibfnamefont {E.~I.}\ \bibnamefont
  {Staroselskaya}}, \bibinfo {author} {\bibfnamefont {J.}~\bibnamefont
  {Venzke}},\ and\ \bibinfo {author} {\bibfnamefont {K.}~\bibnamefont
  {Bartschat}},\ }\bibfield  {title} {\bibinfo {title} {Photoelectron angular
  distributions in bichromatic atomic ionization induced by circularly
  polarized vuv femtosecond pulses},\ }\href
  {https://doi.org/10.1103/PhysRevA.93.033402} {\bibfield  {journal} {\bibinfo
  {journal} {Phys. Rev. A}\ }\textbf {\bibinfo {volume} {93}},\ \bibinfo
  {pages} {033402} (\bibinfo {year} {2016})}\BibitemShut {NoStop}%
\bibitem [{\citenamefont {Gryzlova}\ \emph {et~al.}(2019)\citenamefont
  {Gryzlova}, \citenamefont {Popova}, \citenamefont {Grum-Grzhimailo},
  \citenamefont {Staroselskaya}, \citenamefont {Douguet},\ and\ \citenamefont
  {Bartschat}}]{Gryzlova19}%
  \BibitemOpen
  \bibfield  {author} {\bibinfo {author} {\bibfnamefont {E.~V.}\ \bibnamefont
  {Gryzlova}}, \bibinfo {author} {\bibfnamefont {M.~M.}\ \bibnamefont
  {Popova}}, \bibinfo {author} {\bibfnamefont {A.~N.}\ \bibnamefont
  {Grum-Grzhimailo}}, \bibinfo {author} {\bibfnamefont {E.~I.}\ \bibnamefont
  {Staroselskaya}}, \bibinfo {author} {\bibfnamefont {N.}~\bibnamefont
  {Douguet}},\ and\ \bibinfo {author} {\bibfnamefont {K.}~\bibnamefont
  {Bartschat}},\ }\bibfield  {title} {\bibinfo {title} {Coherent control of the
  photoelectron angular distribution in ionization of neon by a circularly
  polarized bichromatic field in the resonance region},\ }\href@noop {}
  {\bibfield  {journal} {\bibinfo  {journal} {Phys. Rev. A}\ }\textbf {\bibinfo
  {volume} {100}},\ \bibinfo {pages} {063417} (\bibinfo {year}
  {2019})}\BibitemShut {NoStop}%
\bibitem [{\citenamefont {Jašarević}\ \emph {et~al.}(2023)\citenamefont
  {Jašarević}, \citenamefont {Hasović},\ and\ \citenamefont
  {Milošević}}]{Jasarevic2023}%
  \BibitemOpen
  \bibfield  {author} {\bibinfo {author} {\bibfnamefont {A.~S.}\ \bibnamefont
  {Jašarević}}, \bibinfo {author} {\bibfnamefont {E.}~\bibnamefont
  {Hasović}},\ and\ \bibinfo {author} {\bibfnamefont {D.~B.}\ \bibnamefont
  {Milošević}},\ }\bibfield  {title} {\bibinfo {title} {High-order
  above-threshold ionization using a bi-elliptic orthogonal two-color laser
  field with optimal field parameters},\ }\bibfield  {journal} {\bibinfo
  {journal} {Atoms}\ }\textbf {\bibinfo {volume} {11}},\ \href
  {https://doi.org/10.3390/atoms11060091} {10.3390/atoms11060091} (\bibinfo
  {year} {2023})\BibitemShut {NoStop}%
\bibitem [{\citenamefont {Gebre}\ \emph {et~al.}(2024)\citenamefont {Gebre},
  \citenamefont {Walker},\ and\ \citenamefont {Becker}}]{Gebre2024}%
  \BibitemOpen
  \bibfield  {author} {\bibinfo {author} {\bibfnamefont {Y.}~\bibnamefont
  {Gebre}}, \bibinfo {author} {\bibfnamefont {S.}~\bibnamefont {Walker}},\ and\
  \bibinfo {author} {\bibfnamefont {A.}~\bibnamefont {Becker}},\ }\bibfield
  {title} {\bibinfo {title} {Photoelectron spectra in circularly and
  elliptically polarized laser pulses},\ }\href
  {https://doi.org/10.1103/PhysRevA.109.023120} {\bibfield  {journal} {\bibinfo
   {journal} {Phys. Rev. A}\ }\textbf {\bibinfo {volume} {109}},\ \bibinfo
  {pages} {023120} (\bibinfo {year} {2024})}\BibitemShut {NoStop}%
\bibitem [{\citenamefont {Itatani}\ \emph {et~al.}(2002)\citenamefont
  {Itatani}, \citenamefont {Qu\'er\'e}, \citenamefont {Yudin}, \citenamefont
  {Ivanov}, \citenamefont {Krausz},\ and\ \citenamefont
  {Corkum}}]{Itatani2002}%
  \BibitemOpen
  \bibfield  {author} {\bibinfo {author} {\bibfnamefont {J.}~\bibnamefont
  {Itatani}}, \bibinfo {author} {\bibfnamefont {F.}~\bibnamefont {Qu\'er\'e}},
  \bibinfo {author} {\bibfnamefont {G.~L.}\ \bibnamefont {Yudin}}, \bibinfo
  {author} {\bibfnamefont {M.~Y.}\ \bibnamefont {Ivanov}}, \bibinfo {author}
  {\bibfnamefont {F.}~\bibnamefont {Krausz}},\ and\ \bibinfo {author}
  {\bibfnamefont {P.~B.}\ \bibnamefont {Corkum}},\ }\bibfield  {title}
  {\bibinfo {title} {Attosecond streak camera},\ }\href
  {https://doi.org/10.1103/PhysRevLett.88.173903} {\bibfield  {journal}
  {\bibinfo  {journal} {Phys. Rev. Lett.}\ }\textbf {\bibinfo {volume} {88}},\
  \bibinfo {pages} {173903} (\bibinfo {year} {2002})}\BibitemShut {NoStop}%
\bibitem [{\citenamefont {V\'eniard}\ \emph {et~al.}(1996)\citenamefont
  {V\'eniard}, \citenamefont {Ta\"{\i}eb},\ and\ \citenamefont
  {Maquet}}]{Veniard1996}%
  \BibitemOpen
  \bibfield  {author} {\bibinfo {author} {\bibfnamefont {V.}~\bibnamefont
  {V\'eniard}}, \bibinfo {author} {\bibfnamefont {R.}~\bibnamefont
  {Ta\"{\i}eb}},\ and\ \bibinfo {author} {\bibfnamefont {A.}~\bibnamefont
  {Maquet}},\ }\bibfield  {title} {\bibinfo {title} {Phase dependence of
  ($n+1$)-color ($n>1$) ir-uv photoionization of atoms with higher harmonics},\
  }\href {https://doi.org/10.1103/PhysRevA.54.721} {\bibfield  {journal}
  {\bibinfo  {journal} {Phys. Rev. A}\ }\textbf {\bibinfo {volume} {54}},\
  \bibinfo {pages} {721} (\bibinfo {year} {1996})}\BibitemShut {NoStop}%
\bibitem [{\citenamefont {Paul}\ \emph {et~al.}(2001)\citenamefont {Paul},
  \citenamefont {Toma}, \citenamefont {Breger}, \citenamefont {Mullot},
  \citenamefont {Aug\'{e}}, \citenamefont {Balcou}, \citenamefont {Muller},\
  and\ \citenamefont {Agostini}}]{Paul2001}%
  \BibitemOpen
  \bibfield  {author} {\bibinfo {author} {\bibfnamefont {P.~M.}\ \bibnamefont
  {Paul}}, \bibinfo {author} {\bibfnamefont {E.~S.}\ \bibnamefont {Toma}},
  \bibinfo {author} {\bibfnamefont {P.}~\bibnamefont {Breger}}, \bibinfo
  {author} {\bibfnamefont {G.}~\bibnamefont {Mullot}}, \bibinfo {author}
  {\bibfnamefont {F.}~\bibnamefont {Aug\'{e}}}, \bibinfo {author}
  {\bibfnamefont {P.}~\bibnamefont {Balcou}}, \bibinfo {author} {\bibfnamefont
  {H.~G.}\ \bibnamefont {Muller}},\ and\ \bibinfo {author} {\bibfnamefont
  {P.}~\bibnamefont {Agostini}},\ }\bibfield  {title} {\bibinfo {title}
  {Observation of a train of attosecond pulses from high harmonic generation},\
  }\href {https://doi.org/10.1126/science.1059413} {\bibfield  {journal}
  {\bibinfo  {journal} {Science}\ }\textbf {\bibinfo {volume} {292}},\ \bibinfo
  {pages} {1689} (\bibinfo {year} {2001})},\ \Eprint
  {https://arxiv.org/abs/https://www.science.org/doi/pdf/10.1126/science.1059413}
  {https://www.science.org/doi/pdf/10.1126/science.1059413} \BibitemShut
  {NoStop}%
\bibitem [{\citenamefont {Smirnova}\ \emph {et~al.}(2009)\citenamefont
  {Smirnova}, \citenamefont {Mairesse}, \citenamefont {Patchkovskii},
  \citenamefont {Dudovich}, \citenamefont {Villeneuve}, \citenamefont
  {Corkum},\ and\ \citenamefont {Ivanov}}]{smirnova2009high}%
  \BibitemOpen
  \bibfield  {author} {\bibinfo {author} {\bibfnamefont {O.}~\bibnamefont
  {Smirnova}}, \bibinfo {author} {\bibfnamefont {Y.}~\bibnamefont {Mairesse}},
  \bibinfo {author} {\bibfnamefont {S.}~\bibnamefont {Patchkovskii}}, \bibinfo
  {author} {\bibfnamefont {N.}~\bibnamefont {Dudovich}}, \bibinfo {author}
  {\bibfnamefont {D.}~\bibnamefont {Villeneuve}}, \bibinfo {author}
  {\bibfnamefont {P.}~\bibnamefont {Corkum}},\ and\ \bibinfo {author}
  {\bibfnamefont {M.~Y.}\ \bibnamefont {Ivanov}},\ }\bibfield  {title}
  {\bibinfo {title} {High harmonic interferometry of multi-electron dynamics in
  molecules},\ }\href@noop {} {\bibfield  {journal} {\bibinfo  {journal}
  {Nature}\ }\textbf {\bibinfo {volume} {460}},\ \bibinfo {pages} {972}
  (\bibinfo {year} {2009})}\BibitemShut {NoStop}%
\bibitem [{\citenamefont {Moioli}\ \emph {et~al.}(2025)\citenamefont {Moioli},
  \citenamefont {Popova}, \citenamefont {Hamilton}, \citenamefont {Ertel},
  \citenamefont {Busto}, \citenamefont {Makos}, \citenamefont {Kiselev},
  \citenamefont {Yudin}, \citenamefont {Ahmadi}, \citenamefont {Schr\"oter},
  \citenamefont {Pfeifer}, \citenamefont {Moshammer}, \citenamefont {Gryzlova},
  \citenamefont {Grum-Grzhimailo}, \citenamefont {Bartschat},\ and\
  \citenamefont {Sansone}}]{moioli2025}%
  \BibitemOpen
  \bibfield  {author} {\bibinfo {author} {\bibfnamefont {M.}~\bibnamefont
  {Moioli}}, \bibinfo {author} {\bibfnamefont {M.~M.}\ \bibnamefont {Popova}},
  \bibinfo {author} {\bibfnamefont {K.~R.}\ \bibnamefont {Hamilton}}, \bibinfo
  {author} {\bibfnamefont {D.}~\bibnamefont {Ertel}}, \bibinfo {author}
  {\bibfnamefont {D.}~\bibnamefont {Busto}}, \bibinfo {author} {\bibfnamefont
  {I.}~\bibnamefont {Makos}}, \bibinfo {author} {\bibfnamefont {M.~D.}\
  \bibnamefont {Kiselev}}, \bibinfo {author} {\bibfnamefont {S.~N.}\
  \bibnamefont {Yudin}}, \bibinfo {author} {\bibfnamefont {H.}~\bibnamefont
  {Ahmadi}}, \bibinfo {author} {\bibfnamefont {C.~D.}\ \bibnamefont
  {Schr\"oter}}, \bibinfo {author} {\bibfnamefont {T.}~\bibnamefont {Pfeifer}},
  \bibinfo {author} {\bibfnamefont {R.}~\bibnamefont {Moshammer}}, \bibinfo
  {author} {\bibfnamefont {E.~V.}\ \bibnamefont {Gryzlova}}, \bibinfo {author}
  {\bibfnamefont {A.~N.}\ \bibnamefont {Grum-Grzhimailo}}, \bibinfo {author}
  {\bibfnamefont {K.}~\bibnamefont {Bartschat}},\ and\ \bibinfo {author}
  {\bibfnamefont {G.}~\bibnamefont {Sansone}},\ }\bibfield  {title} {\bibinfo
  {title} {Role of intermediate resonances in attosecond photoelectron
  interferometry in neon},\ }\href
  {https://doi.org/10.1103/PhysRevResearch.7.023034} {\bibfield  {journal}
  {\bibinfo  {journal} {Phys. Rev. Res.}\ }\textbf {\bibinfo {volume} {7}},\
  \bibinfo {pages} {023034} (\bibinfo {year} {2025})}\BibitemShut {NoStop}%
\bibitem [{\citenamefont {You}\ \emph {et~al.}(2020)\citenamefont {You},
  \citenamefont {Ueda}, \citenamefont {Gryzlova}, \citenamefont
  {Grum-Grzhimailo}, \citenamefont {Popova}, \citenamefont {Staroselskaya},
  \citenamefont {Tugs}, \citenamefont {Orimo}, \citenamefont {Sato},
  \citenamefont {Ishikawa}, \citenamefont {Carpeggiani}, \citenamefont
  {Csizmadia}, \citenamefont {F\"ule}, \citenamefont {Sansone}, \citenamefont
  {Maroju}, \citenamefont {D'Elia}, \citenamefont {Mazza}, \citenamefont
  {Meyer}, \citenamefont {Callegari}, \citenamefont {Di~Fraia}, \citenamefont
  {Plekan}, \citenamefont {Richter}, \citenamefont {Giannessi}, \citenamefont
  {Allaria}, \citenamefont {De~Ninno}, \citenamefont {Trov\`o}, \citenamefont
  {Badano}, \citenamefont {Diviacco}, \citenamefont {Gaio}, \citenamefont
  {Gauthier}, \citenamefont {Mirian}, \citenamefont {Penco}, \citenamefont
  {Ribi\ifmmode~\check{c}\else \v{c}\fi{}}, \citenamefont {Spampinati},
  \citenamefont {Spezzani},\ and\ \citenamefont {Prince}}]{you2020}%
  \BibitemOpen
  \bibfield  {author} {\bibinfo {author} {\bibfnamefont {D.}~\bibnamefont
  {You}}, \bibinfo {author} {\bibfnamefont {K.}~\bibnamefont {Ueda}}, \bibinfo
  {author} {\bibfnamefont {E.~V.}\ \bibnamefont {Gryzlova}}, \bibinfo {author}
  {\bibfnamefont {A.~N.}\ \bibnamefont {Grum-Grzhimailo}}, \bibinfo {author}
  {\bibfnamefont {M.~M.}\ \bibnamefont {Popova}}, \bibinfo {author}
  {\bibfnamefont {E.~I.}\ \bibnamefont {Staroselskaya}}, \bibinfo {author}
  {\bibfnamefont {O.}~\bibnamefont {Tugs}}, \bibinfo {author} {\bibfnamefont
  {Y.}~\bibnamefont {Orimo}}, \bibinfo {author} {\bibfnamefont
  {T.}~\bibnamefont {Sato}}, \bibinfo {author} {\bibfnamefont {K.~L.}\
  \bibnamefont {Ishikawa}}, \bibinfo {author} {\bibfnamefont {P.~A.}\
  \bibnamefont {Carpeggiani}}, \bibinfo {author} {\bibfnamefont
  {T.}~\bibnamefont {Csizmadia}}, \bibinfo {author} {\bibfnamefont
  {M.}~\bibnamefont {F\"ule}}, \bibinfo {author} {\bibfnamefont
  {G.}~\bibnamefont {Sansone}}, \bibinfo {author} {\bibfnamefont {P.~K.}\
  \bibnamefont {Maroju}}, \bibinfo {author} {\bibfnamefont {A.}~\bibnamefont
  {D'Elia}}, \bibinfo {author} {\bibfnamefont {T.}~\bibnamefont {Mazza}},
  \bibinfo {author} {\bibfnamefont {M.}~\bibnamefont {Meyer}}, \bibinfo
  {author} {\bibfnamefont {C.}~\bibnamefont {Callegari}}, \bibinfo {author}
  {\bibfnamefont {M.}~\bibnamefont {Di~Fraia}}, \bibinfo {author}
  {\bibfnamefont {O.}~\bibnamefont {Plekan}}, \bibinfo {author} {\bibfnamefont
  {R.}~\bibnamefont {Richter}}, \bibinfo {author} {\bibfnamefont
  {L.}~\bibnamefont {Giannessi}}, \bibinfo {author} {\bibfnamefont
  {E.}~\bibnamefont {Allaria}}, \bibinfo {author} {\bibfnamefont
  {G.}~\bibnamefont {De~Ninno}}, \bibinfo {author} {\bibfnamefont
  {M.}~\bibnamefont {Trov\`o}}, \bibinfo {author} {\bibfnamefont
  {L.}~\bibnamefont {Badano}}, \bibinfo {author} {\bibfnamefont
  {B.}~\bibnamefont {Diviacco}}, \bibinfo {author} {\bibfnamefont
  {G.}~\bibnamefont {Gaio}}, \bibinfo {author} {\bibfnamefont {D.}~\bibnamefont
  {Gauthier}}, \bibinfo {author} {\bibfnamefont {N.}~\bibnamefont {Mirian}},
  \bibinfo {author} {\bibfnamefont {G.}~\bibnamefont {Penco}}, \bibinfo
  {author} {\bibfnamefont {P.~c. v.~R.}\ \bibnamefont
  {Ribi\ifmmode~\check{c}\else \v{c}\fi{}}}, \bibinfo {author} {\bibfnamefont
  {S.}~\bibnamefont {Spampinati}}, \bibinfo {author} {\bibfnamefont
  {C.}~\bibnamefont {Spezzani}},\ and\ \bibinfo {author} {\bibfnamefont
  {K.~C.}\ \bibnamefont {Prince}},\ }\bibfield  {title} {\bibinfo {title} {New
  method for measuring angle-resolved phases in photoemission},\ }\href
  {https://doi.org/10.1103/PhysRevX.10.031070} {\bibfield  {journal} {\bibinfo
  {journal} {Phys. Rev. X}\ }\textbf {\bibinfo {volume} {10}},\ \bibinfo
  {pages} {031070} (\bibinfo {year} {2020})}\BibitemShut {NoStop}%
\bibitem [{\citenamefont {Bharti}\ \emph {et~al.}(2021)\citenamefont {Bharti},
  \citenamefont {Atri-Schuller}, \citenamefont {Menning}, \citenamefont
  {Hamilton}, \citenamefont {Moshammer}, \citenamefont {Pfeifer}, \citenamefont
  {Douguet}, \citenamefont {Bartschat},\ and\ \citenamefont
  {Harth}}]{Bharti2021}%
  \BibitemOpen
  \bibfield  {author} {\bibinfo {author} {\bibfnamefont {D.}~\bibnamefont
  {Bharti}}, \bibinfo {author} {\bibfnamefont {D.}~\bibnamefont
  {Atri-Schuller}}, \bibinfo {author} {\bibfnamefont {G.}~\bibnamefont
  {Menning}}, \bibinfo {author} {\bibfnamefont {K.~R.}\ \bibnamefont
  {Hamilton}}, \bibinfo {author} {\bibfnamefont {R.}~\bibnamefont {Moshammer}},
  \bibinfo {author} {\bibfnamefont {T.}~\bibnamefont {Pfeifer}}, \bibinfo
  {author} {\bibfnamefont {N.}~\bibnamefont {Douguet}}, \bibinfo {author}
  {\bibfnamefont {K.}~\bibnamefont {Bartschat}},\ and\ \bibinfo {author}
  {\bibfnamefont {A.}~\bibnamefont {Harth}},\ }\bibfield  {title} {\bibinfo
  {title} {Decomposition of the transition phase in multi-sideband schemes for
  reconstruction of attosecond beating by interference of two-photon
  transitions},\ }\href {https://doi.org/10.1103/PhysRevA.103.022834}
  {\bibfield  {journal} {\bibinfo  {journal} {Phys. Rev. A}\ }\textbf {\bibinfo
  {volume} {103}},\ \bibinfo {pages} {022834} (\bibinfo {year}
  {2021})}\BibitemShut {NoStop}%
\bibitem [{\citenamefont {Fan}\ \emph {et~al.}(2015)\citenamefont {Fan},
  \citenamefont {Grychtol}, \citenamefont {Knut}, \citenamefont
  {Hernández-García}, \citenamefont {Hickstein}, \citenamefont {Zusin},
  \citenamefont {Gentry}, \citenamefont {Dollar}, \citenamefont {Mancuso},
  \citenamefont {Hogle}, \citenamefont {Kfir}, \citenamefont {Legut},
  \citenamefont {Carva}, \citenamefont {Ellis}, \citenamefont {Dorney},
  \citenamefont {Chen}, \citenamefont {Shpyrko}, \citenamefont {Fullerton},
  \citenamefont {Cohen}, \citenamefont {Oppeneer}, \citenamefont {Milošević},
  \citenamefont {Becker}, \citenamefont {Jaroń-Becker}, \citenamefont
  {Popmintchev}, \citenamefont {Murnane},\ and\ \citenamefont {Kapteyn}}]{Fan}%
  \BibitemOpen
  \bibfield  {author} {\bibinfo {author} {\bibfnamefont {T.}~\bibnamefont
  {Fan}}, \bibinfo {author} {\bibfnamefont {P.}~\bibnamefont {Grychtol}},
  \bibinfo {author} {\bibfnamefont {R.}~\bibnamefont {Knut}}, \bibinfo {author}
  {\bibfnamefont {C.}~\bibnamefont {Hernández-García}}, \bibinfo {author}
  {\bibfnamefont {D.~D.}\ \bibnamefont {Hickstein}}, \bibinfo {author}
  {\bibfnamefont {D.}~\bibnamefont {Zusin}}, \bibinfo {author} {\bibfnamefont
  {C.}~\bibnamefont {Gentry}}, \bibinfo {author} {\bibfnamefont {F.~J.}\
  \bibnamefont {Dollar}}, \bibinfo {author} {\bibfnamefont {C.~A.}\
  \bibnamefont {Mancuso}}, \bibinfo {author} {\bibfnamefont {C.~W.}\
  \bibnamefont {Hogle}}, \bibinfo {author} {\bibfnamefont {O.}~\bibnamefont
  {Kfir}}, \bibinfo {author} {\bibfnamefont {D.}~\bibnamefont {Legut}},
  \bibinfo {author} {\bibfnamefont {K.}~\bibnamefont {Carva}}, \bibinfo
  {author} {\bibfnamefont {J.~L.}\ \bibnamefont {Ellis}}, \bibinfo {author}
  {\bibfnamefont {K.~M.}\ \bibnamefont {Dorney}}, \bibinfo {author}
  {\bibfnamefont {C.}~\bibnamefont {Chen}}, \bibinfo {author} {\bibfnamefont
  {O.~G.}\ \bibnamefont {Shpyrko}}, \bibinfo {author} {\bibfnamefont {E.~E.}\
  \bibnamefont {Fullerton}}, \bibinfo {author} {\bibfnamefont {O.}~\bibnamefont
  {Cohen}}, \bibinfo {author} {\bibfnamefont {P.~M.}\ \bibnamefont {Oppeneer}},
  \bibinfo {author} {\bibfnamefont {D.~B.}\ \bibnamefont {Milošević}},
  \bibinfo {author} {\bibfnamefont {A.}~\bibnamefont {Becker}}, \bibinfo
  {author} {\bibfnamefont {A.~A.}\ \bibnamefont {Jaroń-Becker}}, \bibinfo
  {author} {\bibfnamefont {T.}~\bibnamefont {Popmintchev}}, \bibinfo {author}
  {\bibfnamefont {M.~M.}\ \bibnamefont {Murnane}},\ and\ \bibinfo {author}
  {\bibfnamefont {H.~C.}\ \bibnamefont {Kapteyn}},\ }\bibfield  {title}
  {\bibinfo {title} {Bright circularly polarized soft x-ray high harmonics for
  x-ray magnetic circular dichroism},\ }\href
  {https://doi.org/10.1073/pnas.1519666112} {\bibfield  {journal} {\bibinfo
  {journal} {Proceedings of the National Academy of Sciences}\ }\textbf
  {\bibinfo {volume} {112}},\ \bibinfo {pages} {14206} (\bibinfo {year}
  {2015})},\ \Eprint
  {https://arxiv.org/abs/https://www.pnas.org/doi/pdf/10.1073/pnas.1519666112}
  {https://www.pnas.org/doi/pdf/10.1073/pnas.1519666112} \BibitemShut {NoStop}%
\bibitem [{\citenamefont {Barreau}\ \emph {et~al.}(2018)\citenamefont
  {Barreau}, \citenamefont {Veyrinas}, \citenamefont {Gruson}, \citenamefont
  {J.Weber}, \citenamefont {Auguste}, \citenamefont {Hergott}, \citenamefont
  {Lepetit}, \citenamefont {Carré}, \citenamefont {Houver}, \citenamefont
  {Dowek},\ and\ \citenamefont {Salières}}]{Barreau}%
  \BibitemOpen
  \bibfield  {author} {\bibinfo {author} {\bibfnamefont {L.}~\bibnamefont
  {Barreau}}, \bibinfo {author} {\bibfnamefont {K.}~\bibnamefont {Veyrinas}},
  \bibinfo {author} {\bibfnamefont {V.}~\bibnamefont {Gruson}}, \bibinfo
  {author} {\bibfnamefont {S.}~\bibnamefont {J.Weber}}, \bibinfo {author}
  {\bibfnamefont {T.}~\bibnamefont {Auguste}}, \bibinfo {author} {\bibfnamefont
  {J.-F.}\ \bibnamefont {Hergott}}, \bibinfo {author} {\bibfnamefont
  {F.}~\bibnamefont {Lepetit}}, \bibinfo {author} {\bibfnamefont
  {B.}~\bibnamefont {Carré}}, \bibinfo {author} {\bibfnamefont {J.-C.}\
  \bibnamefont {Houver}}, \bibinfo {author} {\bibfnamefont {D.}~\bibnamefont
  {Dowek}},\ and\ \bibinfo {author} {\bibfnamefont {P.}~\bibnamefont
  {Salières}},\ }\href@noop {} {\bibfield  {journal} {\bibinfo  {journal}
  {Nat. Commun.}\ }\textbf {\bibinfo {volume} {9}},\ \bibinfo {pages} {4727}
  (\bibinfo {year} {2018})}\BibitemShut {NoStop}%
\bibitem [{\citenamefont {Villeneuve}\ \emph {et~al.}(2017)\citenamefont
  {Villeneuve}, \citenamefont {Hockett}, \citenamefont {Vrakking},\ and\
  \citenamefont {Niikura}}]{villeneuve2017}%
  \BibitemOpen
  \bibfield  {author} {\bibinfo {author} {\bibfnamefont {D.}~\bibnamefont
  {Villeneuve}}, \bibinfo {author} {\bibfnamefont {P.}~\bibnamefont {Hockett}},
  \bibinfo {author} {\bibfnamefont {M.}~\bibnamefont {Vrakking}},\ and\
  \bibinfo {author} {\bibfnamefont {H.}~\bibnamefont {Niikura}},\ }\bibfield
  {title} {\bibinfo {title} {Coherent imaging of an attosecond electron wave
  packet},\ }\href@noop {} {\bibfield  {journal} {\bibinfo  {journal}
  {Science}\ }\textbf {\bibinfo {volume} {356}},\ \bibinfo {pages} {1150}
  (\bibinfo {year} {2017})}\BibitemShut {NoStop}%
\bibitem [{\citenamefont {Donsa}\ \emph {et~al.}(2019)\citenamefont {Donsa},
  \citenamefont {Douguet}, \citenamefont {Burgd\"orfer}, \citenamefont
  {B\ifmmode~\check{r}\else \v{r}\fi{}ezinov\'a},\ and\ \citenamefont
  {Argenti}}]{Donsa2019}%
  \BibitemOpen
  \bibfield  {author} {\bibinfo {author} {\bibfnamefont {S.}~\bibnamefont
  {Donsa}}, \bibinfo {author} {\bibfnamefont {N.}~\bibnamefont {Douguet}},
  \bibinfo {author} {\bibfnamefont {J.}~\bibnamefont {Burgd\"orfer}}, \bibinfo
  {author} {\bibfnamefont {I.}~\bibnamefont {B\ifmmode~\check{r}\else
  \v{r}\fi{}ezinov\'a}},\ and\ \bibinfo {author} {\bibfnamefont
  {L.}~\bibnamefont {Argenti}},\ }\bibfield  {title} {\bibinfo {title}
  {Circular holographic ionization-phase meter},\ }\href
  {https://doi.org/10.1103/PhysRevLett.123.133203} {\bibfield  {journal}
  {\bibinfo  {journal} {Phys. Rev. Lett.}\ }\textbf {\bibinfo {volume} {123}},\
  \bibinfo {pages} {133203} (\bibinfo {year} {2019})}\BibitemShut {NoStop}%
\bibitem [{\citenamefont {Barth}\ and\ \citenamefont
  {Smirnova}(2013)}]{Barth13}%
  \BibitemOpen
  \bibfield  {author} {\bibinfo {author} {\bibfnamefont {I.}~\bibnamefont
  {Barth}}\ and\ \bibinfo {author} {\bibfnamefont {O.}~\bibnamefont
  {Smirnova}},\ }\bibfield  {title} {\bibinfo {title} {Spin-polarized electrons
  produced by strong-field ionization},\ }\href@noop {} {\bibfield  {journal}
  {\bibinfo  {journal} {Phys. Rev. A.}\ }\textbf {\bibinfo {volume} {88}},\
  \bibinfo {pages} {013401} (\bibinfo {year} {2013})}\BibitemShut {NoStop}%
\bibitem [{\citenamefont {Hartung}\ \emph {et~al.}(2016)\citenamefont
  {Hartung}, \citenamefont {Morales}, \citenamefont {Kunitski}, \citenamefont
  {Henrichs}, \citenamefont {Laucke}, \citenamefont {Richter}, \citenamefont
  {Jahnke}, \citenamefont {Kalinin}, \citenamefont {Schöffler}, \citenamefont
  {Schmidt}, \citenamefont {Ivanov}, \citenamefont {Smirnova},\ and\
  \citenamefont {Reinhard}}]{Hartung16}%
  \BibitemOpen
  \bibfield  {author} {\bibinfo {author} {\bibfnamefont {A.}~\bibnamefont
  {Hartung}}, \bibinfo {author} {\bibfnamefont {F.}~\bibnamefont {Morales}},
  \bibinfo {author} {\bibfnamefont {M.}~\bibnamefont {Kunitski}}, \bibinfo
  {author} {\bibfnamefont {K.}~\bibnamefont {Henrichs}}, \bibinfo {author}
  {\bibfnamefont {A.}~\bibnamefont {Laucke}}, \bibinfo {author} {\bibfnamefont
  {M.}~\bibnamefont {Richter}}, \bibinfo {author} {\bibfnamefont
  {T.}~\bibnamefont {Jahnke}}, \bibinfo {author} {\bibfnamefont
  {A.}~\bibnamefont {Kalinin}}, \bibinfo {author} {\bibfnamefont
  {M.}~\bibnamefont {Schöffler}}, \bibinfo {author} {\bibfnamefont {L.~P.~H.}\
  \bibnamefont {Schmidt}}, \bibinfo {author} {\bibfnamefont {M.}~\bibnamefont
  {Ivanov}}, \bibinfo {author} {\bibfnamefont {O.}~\bibnamefont {Smirnova}},\
  and\ \bibinfo {author} {\bibfnamefont {D.}~\bibnamefont {Reinhard}},\
  }\bibfield  {title} {\bibinfo {title} {Electron spin polarization in
  strong-field ionization of xenon atoms},\ }\href@noop {} {\bibfield
  {journal} {\bibinfo  {journal} {Nat. Photonics}\ }\textbf {\bibinfo {volume}
  {10}},\ \bibinfo {pages} {526} (\bibinfo {year} {2016})}\BibitemShut
  {NoStop}%
\bibitem [{\citenamefont {Milo\ifmmode \check{s}\else
  \v{s}\fi{}evi\ifmmode~\acute{c}\else \'{c}\fi{}}(2016)}]{Milocevic2016}%
  \BibitemOpen
  \bibfield  {author} {\bibinfo {author} {\bibfnamefont {D.~B.}\ \bibnamefont
  {Milo\ifmmode \check{s}\else \v{s}\fi{}evi\ifmmode~\acute{c}\else
  \'{c}\fi{}}},\ }\bibfield  {title} {\bibinfo {title} {Possibility of
  introducing spin into attoscience with spin-polarized electrons produced by a
  bichromatic circularly polarized laser field},\ }\href
  {https://doi.org/10.1103/PhysRevA.93.051402} {\bibfield  {journal} {\bibinfo
  {journal} {Phys. Rev. A}\ }\textbf {\bibinfo {volume} {93}},\ \bibinfo
  {pages} {051402} (\bibinfo {year} {2016})}\BibitemShut {NoStop}%
\bibitem [{\citenamefont {Gryzlova}\ \emph {et~al.}(2020)\citenamefont
  {Gryzlova}, \citenamefont {Popova},\ and\ \citenamefont
  {Grum-Grzhimailo}}]{Gryzlova2020_2}%
  \BibitemOpen
  \bibfield  {author} {\bibinfo {author} {\bibfnamefont {E.~V.}\ \bibnamefont
  {Gryzlova}}, \bibinfo {author} {\bibfnamefont {M.~M.}\ \bibnamefont
  {Popova}},\ and\ \bibinfo {author} {\bibfnamefont {A.~N.}\ \bibnamefont
  {Grum-Grzhimailo}},\ }\bibfield  {title} {\bibinfo {title} {Spin polarization
  of photoelectrons in bichromatic extreme-ultraviolet atomic ionization},\
  }\href {https://doi.org/10.1103/PhysRevA.102.053116} {\bibfield  {journal}
  {\bibinfo  {journal} {Phys. Rev. A}\ }\textbf {\bibinfo {volume} {102}},\
  \bibinfo {pages} {053116} (\bibinfo {year} {2020})}\BibitemShut {NoStop}%
\bibitem [{\citenamefont {Goetz}\ \emph {et~al.}(2019)\citenamefont {Goetz},
  \citenamefont {Koch},\ and\ \citenamefont {Greenman}}]{Goetz2019}%
  \BibitemOpen
  \bibfield  {author} {\bibinfo {author} {\bibfnamefont {R.~E.}\ \bibnamefont
  {Goetz}}, \bibinfo {author} {\bibfnamefont {C.~P.}\ \bibnamefont {Koch}},\
  and\ \bibinfo {author} {\bibfnamefont {L.}~\bibnamefont {Greenman}},\
  }\bibfield  {title} {\bibinfo {title} {Perfect control of photoelectron
  anisotropy for randomly oriented ensembles of molecules by xuv rempi and
  polarization shaping},\ }\href {https://doi.org/10.1063/1.5111362} {\bibfield
   {journal} {\bibinfo  {journal} {The Journal of Chemical Physics}\ }\textbf
  {\bibinfo {volume} {151}},\ \bibinfo {pages} {074106} (\bibinfo {year}
  {2019})}\BibitemShut {NoStop}%
\bibitem [{\citenamefont {Milošević}(2018)}]{atoms6040061}%
  \BibitemOpen
  \bibfield  {author} {\bibinfo {author} {\bibfnamefont {D.~B.}\ \bibnamefont
  {Milošević}},\ }\bibfield  {title} {\bibinfo {title} {Atomic and molecular
  processes in a strong bicircular laser field},\ }\href
  {https://www.mdpi.com/2218-2004/6/4/61} {\bibfield  {journal} {\bibinfo
  {journal} {Atoms}\ }\textbf {\bibinfo {volume} {6}} (\bibinfo {year}
  {2018})}\BibitemShut {NoStop}%
\bibitem [{\citenamefont {Popova}\ \emph {et~al.}(2022)\citenamefont {Popova},
  \citenamefont {Gryzlova}, \citenamefont {Kiselev},\ and\ \citenamefont
  {Grum-Grzhimailo}}]{Popova2022}%
  \BibitemOpen
  \bibfield  {author} {\bibinfo {author} {\bibfnamefont {M.~M.}\ \bibnamefont
  {Popova}}, \bibinfo {author} {\bibfnamefont {E.~V.}\ \bibnamefont
  {Gryzlova}}, \bibinfo {author} {\bibfnamefont {M.~D.}\ \bibnamefont
  {Kiselev}},\ and\ \bibinfo {author} {\bibfnamefont {A.~N.}\ \bibnamefont
  {Grum-Grzhimailo}},\ }\bibfield  {title} {\bibinfo {title} {Ionization of
  atoms by a bichromatic fields of $\omega+2\omega$ multiple frequencies with
  arbitrary polarization},\ }\href
  {https://doi.org/doi.org/10.1134/S1063776122070044} {\bibfield  {journal}
  {\bibinfo  {journal} {Journal of Experimental and Theoretical Physics}\
  }\textbf {\bibinfo {volume} {135}},\ \bibinfo {pages} {58–72} (\bibinfo
  {year} {2022})}\BibitemShut {NoStop}%
\bibitem [{\citenamefont {Hockett}(2017)}]{Hockett2017}%
  \BibitemOpen
  \bibfield  {author} {\bibinfo {author} {\bibfnamefont {P.}~\bibnamefont
  {Hockett}},\ }\bibfield  {title} {\bibinfo {title} {Angle-resolved {RABBITT}:
  theory and numerics},\ }\href {https://doi.org/10.1088/1361-6455/aa7887}
  {\bibfield  {journal} {\bibinfo  {journal} {Journal of Physics B: Atomic,
  Molecular and Optical Physics}\ }\textbf {\bibinfo {volume} {50}},\ \bibinfo
  {pages} {154002} (\bibinfo {year} {2017})}\BibitemShut {NoStop}%
\bibitem [{\citenamefont {Boll}\ and\ \citenamefont {Fojón}(2017)}]{Boll2017}%
  \BibitemOpen
  \bibfield  {author} {\bibinfo {author} {\bibfnamefont {D.~I.~R.}\
  \bibnamefont {Boll}}\ and\ \bibinfo {author} {\bibfnamefont {O.~A.}\
  \bibnamefont {Fojón}},\ }\bibfield  {title} {\bibinfo {title} {Attosecond
  polarization control in atomic {RABBITT}-like experiments assisted by a
  circularly polarized laser},\ }\href@noop {} {\bibfield  {journal} {\bibinfo
  {journal} {J. Phys. B}\ }\textbf {\bibinfo {volume} {50}} (\bibinfo {year}
  {2017})}\BibitemShut {NoStop}%
\bibitem [{\citenamefont {Kheifets}\ and\ \citenamefont
  {Xu}(2023)}]{Kheifets2023}%
  \BibitemOpen
  \bibfield  {author} {\bibinfo {author} {\bibfnamefont {A.~S.}\ \bibnamefont
  {Kheifets}}\ and\ \bibinfo {author} {\bibfnamefont {Z.}~\bibnamefont {Xu}},\
  }\bibfield  {title} {\bibinfo {title} {Polarization control of {RABBITT} in
  noble gas atoms},\ }\href {https://doi.org/10.1088/1361-6455/ace574}
  {\bibfield  {journal} {\bibinfo  {journal} {Journal of Physics B: Atomic,
  Molecular and Optical Physics}\ }\textbf {\bibinfo {volume} {56}},\ \bibinfo
  {pages} {155601} (\bibinfo {year} {2023})}\BibitemShut {NoStop}%
\bibitem [{\citenamefont {Popova}\ \emph {et~al.}(2025)\citenamefont {Popova},
  \citenamefont {Gryzlova}, \citenamefont {Yudin},\ and\ \citenamefont
  {Grum-Grzhimailo}}]{popova2025}%
  \BibitemOpen
  \bibfield  {author} {\bibinfo {author} {\bibfnamefont {M.~M.}\ \bibnamefont
  {Popova}}, \bibinfo {author} {\bibfnamefont {E.~V.}\ \bibnamefont
  {Gryzlova}}, \bibinfo {author} {\bibfnamefont {S.~N.}\ \bibnamefont
  {Yudin}},\ and\ \bibinfo {author} {\bibfnamefont {A.~N.}\ \bibnamefont
  {Grum-Grzhimailo}},\ }\bibfield  {title} {\bibinfo {title} {Advantages of
  polarization control in rabbitt},\ }\href
  {https://doi.org/10.1103/PhysRevA.111.033105} {\bibfield  {journal} {\bibinfo
   {journal} {Phys. Rev. A}\ }\textbf {\bibinfo {volume} {111}},\ \bibinfo
  {pages} {033105} (\bibinfo {year} {2025})}\BibitemShut {NoStop}%
\bibitem [{\citenamefont {Bray}\ \emph {et~al.}(2018)\citenamefont {Bray},
  \citenamefont {Naseem},\ and\ \citenamefont {Kheifets}}]{Bray2018}%
  \BibitemOpen
  \bibfield  {author} {\bibinfo {author} {\bibfnamefont {A.~W.}\ \bibnamefont
  {Bray}}, \bibinfo {author} {\bibfnamefont {F.}~\bibnamefont {Naseem}},\ and\
  \bibinfo {author} {\bibfnamefont {A.~S.}\ \bibnamefont {Kheifets}},\
  }\bibfield  {title} {\bibinfo {title} {Simulation of angular-resolved rabbitt
  measurements in noble-gas atoms},\ }\href
  {https://doi.org/10.1103/PhysRevA.97.063404} {\bibfield  {journal} {\bibinfo
  {journal} {Phys. Rev. A}\ }\textbf {\bibinfo {volume} {97}},\ \bibinfo
  {pages} {063404} (\bibinfo {year} {2018})}\BibitemShut {NoStop}%
\bibitem [{\citenamefont {Ocello}\ \emph {et~al.}(2025)\citenamefont {Ocello},
  \citenamefont {López}, \citenamefont {Barlari},\ and\ \citenamefont
  {Arbó}}]{ocello2025}%
  \BibitemOpen
  \bibfield  {author} {\bibinfo {author} {\bibfnamefont {M.~L.}\ \bibnamefont
  {Ocello}}, \bibinfo {author} {\bibfnamefont {S.~D.}\ \bibnamefont {López}},
  \bibinfo {author} {\bibfnamefont {M.}~\bibnamefont {Barlari}},\ and\ \bibinfo
  {author} {\bibfnamefont {D.~G.}\ \bibnamefont {Arbó}},\ }\bibfield  {title}
  {\bibinfo {title} {Time-dependent theory of electron emission perpendicular
  to laser polarization for reconstruction of attosecond harmonic beating by
  interference of multiphoton transitions},\ }\bibfield  {journal} {\bibinfo
  {journal} {Atoms}\ }\textbf {\bibinfo {volume} {13}},\ \href
  {https://doi.org/10.3390/atoms13120099} {10.3390/atoms13120099} (\bibinfo
  {year} {2025})\BibitemShut {NoStop}%
\bibitem [{\citenamefont {Serov}\ \emph {et~al.}(2026)\citenamefont {Serov},
  \citenamefont {Ji}, \citenamefont {Han}, \citenamefont {Ueda}, \citenamefont
  {W\"orner},\ and\ \citenamefont {Kheifets}}]{Serov2026}%
  \BibitemOpen
  \bibfield  {author} {\bibinfo {author} {\bibfnamefont {V.~V.}\ \bibnamefont
  {Serov}}, \bibinfo {author} {\bibfnamefont {J.-B.}\ \bibnamefont {Ji}},
  \bibinfo {author} {\bibfnamefont {M.}~\bibnamefont {Han}}, \bibinfo {author}
  {\bibfnamefont {K.}~\bibnamefont {Ueda}}, \bibinfo {author} {\bibfnamefont
  {H.~J.}\ \bibnamefont {W\"orner}},\ and\ \bibinfo {author} {\bibfnamefont
  {A.~S.}\ \bibnamefont {Kheifets}},\ }\bibfield  {title} {\bibinfo {title}
  {Circular rabbitt goes under threshold: A sensitive probe of discrete
  excitations in noble gas atoms},\ }\href {https://doi.org/10.1103/z3k5-wqg9}
  {\bibfield  {journal} {\bibinfo  {journal} {Phys. Rev. Lett.}\ }\textbf
  {\bibinfo {volume} {136}},\ \bibinfo {pages} {083202} (\bibinfo {year}
  {2026})}\BibitemShut {NoStop}%
\bibitem [{\citenamefont {Popova}\ \emph {et~al.}(2023)\citenamefont {Popova},
  \citenamefont {Yudin}, \citenamefont {Gryzlova}, \citenamefont {Kiselev},\
  and\ \citenamefont {Grum-Grzhimailo}}]{popova2023}%
  \BibitemOpen
  \bibfield  {author} {\bibinfo {author} {\bibfnamefont {M.}~\bibnamefont
  {Popova}}, \bibinfo {author} {\bibfnamefont {S.}~\bibnamefont {Yudin}},
  \bibinfo {author} {\bibfnamefont {E.}~\bibnamefont {Gryzlova}}, \bibinfo
  {author} {\bibfnamefont {M.}~\bibnamefont {Kiselev}},\ and\ \bibinfo {author}
  {\bibfnamefont {A.}~\bibnamefont {Grum-Grzhimailo}},\ }\bibfield  {title}
  {\bibinfo {title} {Attosecond interferometry involving discrete states},\
  }\href@noop {} {\bibfield  {journal} {\bibinfo  {journal} {Journal of
  Experimental and Theoretical Physics}\ }\textbf {\bibinfo {volume} {136}},\
  \bibinfo {pages} {259} (\bibinfo {year} {2023})}\BibitemShut {NoStop}%
\bibitem [{\citenamefont {Gordon}(1929)}]{Gordon1929}%
  \BibitemOpen
  \bibfield  {author} {\bibinfo {author} {\bibfnamefont {W.}~\bibnamefont
  {Gordon}},\ }\href@noop {} {\bibfield  {journal} {\bibinfo  {journal} {Ann.
  Phys. (Leipzig)}\ ,\ \bibinfo {pages} {1031}} (\bibinfo {year}
  {1929})}\BibitemShut {NoStop}%
\bibitem [{\citenamefont {V\'eniard}\ and\ \citenamefont
  {Piraux}(1990)}]{Veniard1989}%
  \BibitemOpen
  \bibfield  {author} {\bibinfo {author} {\bibfnamefont {V.}~\bibnamefont
  {V\'eniard}}\ and\ \bibinfo {author} {\bibfnamefont {B.}~\bibnamefont
  {Piraux}},\ }\bibfield  {title} {\bibinfo {title} {Continuum-continuum dipole
  transitions in femtosecond-laser-pulse excitation of atomic hydrogen},\
  }\href {https://doi.org/10.1103/PhysRevA.41.4019} {\bibfield  {journal}
  {\bibinfo  {journal} {Phys. Rev. A}\ }\textbf {\bibinfo {volume} {41}},\
  \bibinfo {pages} {4019} (\bibinfo {year} {1990})}\BibitemShut {NoStop}%
\bibitem [{\citenamefont {Trippenbach}\ \emph {et~al.}(1989)\citenamefont
  {Trippenbach}, \citenamefont {Rzazewski}, \citenamefont {Fedorov},\ and\
  \citenamefont {Kazakov}}]{Trippenbach_1989}%
  \BibitemOpen
  \bibfield  {author} {\bibinfo {author} {\bibfnamefont {M.}~\bibnamefont
  {Trippenbach}}, \bibinfo {author} {\bibfnamefont {K.}~\bibnamefont
  {Rzazewski}}, \bibinfo {author} {\bibfnamefont {M.~V.}\ \bibnamefont
  {Fedorov}},\ and\ \bibinfo {author} {\bibfnamefont {A.~E.}\ \bibnamefont
  {Kazakov}},\ }\bibfield  {title} {\bibinfo {title} {Semiclassical matrix
  elements, essential-states models and perturbation theory of above-threshold
  ionisation},\ }\href {https://doi.org/10.1088/0953-4075/22/8/012} {\bibfield
  {journal} {\bibinfo  {journal} {Journal of Physics B: Atomic, Molecular and
  Optical Physics}\ }\textbf {\bibinfo {volume} {22}},\ \bibinfo {pages} {1193}
  (\bibinfo {year} {1989})}\BibitemShut {NoStop}%
\bibitem [{\citenamefont {Korol}(1994)}]{Korol_1994}%
  \BibitemOpen
  \bibfield  {author} {\bibinfo {author} {\bibfnamefont {A.~V.}\ \bibnamefont
  {Korol}},\ }\bibfield  {title} {\bibinfo {title} {General formula for the
  singular part of the free-free dipole matrix element},\ }\href
  {https://doi.org/10.1088/0953-4075/27/6/001} {\bibfield  {journal} {\bibinfo
  {journal} {Journal of Physics B: Atomic, Molecular and Optical Physics}\
  }\textbf {\bibinfo {volume} {27}},\ \bibinfo {pages} {L103} (\bibinfo {year}
  {1994})}\BibitemShut {NoStop}%
\bibitem [{\citenamefont {Bello}\ \emph {et~al.}(2021)\citenamefont {Bello},
  \citenamefont {Lucchese}, \citenamefont {Rescigno},\ and\ \citenamefont
  {McCurdy}}]{bello2021}%
  \BibitemOpen
  \bibfield  {author} {\bibinfo {author} {\bibfnamefont {R.~Y.}\ \bibnamefont
  {Bello}}, \bibinfo {author} {\bibfnamefont {R.~R.}\ \bibnamefont {Lucchese}},
  \bibinfo {author} {\bibfnamefont {T.~N.}\ \bibnamefont {Rescigno}},\ and\
  \bibinfo {author} {\bibfnamefont {C.~W.}\ \bibnamefont {McCurdy}},\
  }\bibfield  {title} {\bibinfo {title} {Correlated variational treatment of
  ionization coupled to nuclear motion: Ultrafast pump and ionizing probe of
  electronic and nuclear dynamics in lih},\ }\href
  {https://doi.org/10.1103/PhysRevResearch.3.013228} {\bibfield  {journal}
  {\bibinfo  {journal} {Phys. Rev. Res.}\ }\textbf {\bibinfo {volume} {3}},\
  \bibinfo {pages} {013228} (\bibinfo {year} {2021})}\BibitemShut {NoStop}%
\bibitem [{\citenamefont {Balashov}\ \emph {et~al.}(2000)\citenamefont
  {Balashov}, \citenamefont {Grum-Grzhimailo},\ and\ \citenamefont
  {Kabachnik}}]{polcor}%
  \BibitemOpen
  \bibfield  {author} {\bibinfo {author} {\bibfnamefont {V.~V.}\ \bibnamefont
  {Balashov}}, \bibinfo {author} {\bibfnamefont {A.~N.}\ \bibnamefont
  {Grum-Grzhimailo}},\ and\ \bibinfo {author} {\bibfnamefont {N.~M.}\
  \bibnamefont {Kabachnik}},\ }\href@noop {} {\emph {\bibinfo {title}
  {Polarization and Correlation Phenomena in Atomic Collisions: A Practical
  Theory Course}}}\ (\bibinfo  {publisher} {Kluwer Academic/Plenum
  Publishers},\ \bibinfo {address} {New York},\ \bibinfo {year}
  {2000})\BibitemShut {NoStop}%
\bibitem [{\citenamefont {{Froese~Fischer}}\ \emph {et~al.}(1997)\citenamefont
  {{Froese~Fischer}}, \citenamefont {Brage},\ and\ \citenamefont
  {J{\"o}nsson}}]{Fischer1997}%
  \BibitemOpen
  \bibfield  {author} {\bibinfo {author} {\bibfnamefont {C.}~\bibnamefont
  {{Froese~Fischer}}}, \bibinfo {author} {\bibfnamefont {T.}~\bibnamefont
  {Brage}},\ and\ \bibinfo {author} {\bibfnamefont {P.}~\bibnamefont
  {J{\"o}nsson}},\ }\href {https://doi.org/10.1201/9781315139982} {\emph
  {\bibinfo {title} {Computational Atomic Structure. An {MCHF} Approach}}}\
  (\bibinfo  {publisher} {Bristol, Institute of Physics Publishing},\ \bibinfo
  {year} {1997})\BibitemShut {NoStop}%
\bibitem [{\citenamefont {Mercouris}\ \emph {et~al.}(1994)\citenamefont
  {Mercouris}, \citenamefont {Komninos}, \citenamefont {Dionissopoulou},\ and\
  \citenamefont {Nicolaides}}]{Mercouris1994}%
  \BibitemOpen
  \bibfield  {author} {\bibinfo {author} {\bibfnamefont {T.}~\bibnamefont
  {Mercouris}}, \bibinfo {author} {\bibfnamefont {Y.}~\bibnamefont {Komninos}},
  \bibinfo {author} {\bibfnamefont {S.}~\bibnamefont {Dionissopoulou}},\ and\
  \bibinfo {author} {\bibfnamefont {C.~A.}\ \bibnamefont {Nicolaides}},\
  }\bibfield  {title} {\bibinfo {title} {Computation of strong-field
  multiphoton processes in polyelectronic atoms: State-specific method and
  applications to h and ${\mathrm{li}}^{\mathrm{\ensuremath{-}}}$},\ }\href
  {https://doi.org/10.1103/PhysRevA.50.4109} {\bibfield  {journal} {\bibinfo
  {journal} {Phys. Rev. A}\ }\textbf {\bibinfo {volume} {50}},\ \bibinfo
  {pages} {4109} (\bibinfo {year} {1994})}\BibitemShut {NoStop}%
\bibitem [{\citenamefont {Novikov}\ and\ \citenamefont
  {Hopersky}(2011)}]{Novikov2011}%
  \BibitemOpen
  \bibfield  {author} {\bibinfo {author} {\bibfnamefont {S.~A.}\ \bibnamefont
  {Novikov}}\ and\ \bibinfo {author} {\bibfnamefont {A.~N.}\ \bibnamefont
  {Hopersky}},\ }\bibfield  {title} {\bibinfo {title} {Free{\textendash}free
  matrix elements for a many-electron atom},\ }\href
  {https://doi.org/10.1088/0953-4075/44/23/235001} {\ \textbf {\bibinfo
  {volume} {44}},\ \bibinfo {pages} {235001} (\bibinfo {year}
  {2011})}\BibitemShut {NoStop}%
\bibitem [{\citenamefont {G.A.~Korn}(1968)}]{MATHEMATICALHANDBOOK}%
  \BibitemOpen
  \bibfield  {author} {\bibinfo {author} {\bibfnamefont {T.~K.}\ \bibnamefont
  {G.A.~Korn}},\ }\href@noop {} {\emph {\bibinfo {title} {MATHEMATICAL
  HANDBOOK, second, extended and revised edition}}}\ (\bibinfo  {publisher}
  {McGraw-Hill Book Company},\ \bibinfo {address} {New York, San Francisco,
  Toronto, London, Sydney},\ \bibinfo {year} {1968})\BibitemShut {NoStop}%
\bibitem [{\citenamefont {Baranova}\ \emph {et~al.}(1990)\citenamefont
  {Baranova}, \citenamefont {Zel'dovich}, \citenamefont {Chudinov},\ and\
  \citenamefont {Shul'ginov}}]{Baranova1990}%
  \BibitemOpen
  \bibfield  {author} {\bibinfo {author} {\bibfnamefont {N.~B.}\ \bibnamefont
  {Baranova}}, \bibinfo {author} {\bibfnamefont {B.~Y.}\ \bibnamefont
  {Zel'dovich}}, \bibinfo {author} {\bibfnamefont {A.~N.}\ \bibnamefont
  {Chudinov}},\ and\ \bibinfo {author} {\bibfnamefont {A.~A.}\ \bibnamefont
  {Shul'ginov}},\ }\bibfield  {title} {\bibinfo {title} {Theory and observation
  of polar asymmetry of photoionization in a field with $<{E}^3>\neq 0$},\
  }\href@noop {} {\bibfield  {journal} {\bibinfo  {journal} {Zh. Eksp. Teor.
  Fiz.}\ }\textbf {\bibinfo {volume} {71}},\ \bibinfo {pages} {1857} (\bibinfo
  {year} {1990})}\BibitemShut {NoStop}%
\bibitem [{\citenamefont {Busto}\ \emph {et~al.}(2019)\citenamefont {Busto},
  \citenamefont {Vinbladh}, \citenamefont {Zhong}, \citenamefont {Isinger},
  \citenamefont {Nandi}, \citenamefont {Maclot}, \citenamefont {Johnsson},
  \citenamefont {Gisselbrecht}, \citenamefont {L'Huillier}, \citenamefont
  {Lindroth},\ and\ \citenamefont {Dahlstr\"om}}]{Busto2016}%
  \BibitemOpen
  \bibfield  {author} {\bibinfo {author} {\bibfnamefont {D.}~\bibnamefont
  {Busto}}, \bibinfo {author} {\bibfnamefont {J.}~\bibnamefont {Vinbladh}},
  \bibinfo {author} {\bibfnamefont {S.}~\bibnamefont {Zhong}}, \bibinfo
  {author} {\bibfnamefont {M.}~\bibnamefont {Isinger}}, \bibinfo {author}
  {\bibfnamefont {S.}~\bibnamefont {Nandi}}, \bibinfo {author} {\bibfnamefont
  {S.}~\bibnamefont {Maclot}}, \bibinfo {author} {\bibfnamefont
  {P.}~\bibnamefont {Johnsson}}, \bibinfo {author} {\bibfnamefont
  {M.}~\bibnamefont {Gisselbrecht}}, \bibinfo {author} {\bibfnamefont
  {A.}~\bibnamefont {L'Huillier}}, \bibinfo {author} {\bibfnamefont
  {E.}~\bibnamefont {Lindroth}},\ and\ \bibinfo {author} {\bibfnamefont
  {J.~M.}\ \bibnamefont {Dahlstr\"om}},\ }\bibfield  {title} {\bibinfo {title}
  {Fano's propensity rule in angle-resolved attosecond pump-probe
  photoionization},\ }\href {https://doi.org/10.1103/PhysRevLett.123.133201}
  {\bibfield  {journal} {\bibinfo  {journal} {Phys. Rev. Lett.}\ }\textbf
  {\bibinfo {volume} {123}},\ \bibinfo {pages} {133201} (\bibinfo {year}
  {2019})}\BibitemShut {NoStop}%
\end{thebibliography}%

\end{document}